\documentclass[twocolumn,twocolappendix]{aastex631}

\usepackage{amsmath}
\usepackage{booktabs}

\shortauthors{Hatta et al.}
\graphicspath{{./}{figures/}}

\begin{document}

\title{Inversion for Inferring Solar Meridional Circulation: \\
The Case with Constraints on Angular Momentum Transport inside the Sun}

\correspondingauthor{Yoshiki Hatta}
\email{yoshiki.hatta@isee.nagoya-u.ac.jp}


\author[0000-0003-0747-8835]{Yoshiki Hatta}
\affiliation{Institute for Space-Earth Environmental Research, Nagoya University \\
Furo-cho, Chikusa-ku, Nagoya, Aichi 464-8601, Japan}
\affiliation{National Astronomical Observatory of Japan \\
2-21-1 Osawa, Mitaka, Tokyo 181-8588, Japan}

\author[0000-0002-6312-7944]{Hideyuki Hotta}
\affiliation{Institute for Space-Earth Environmental Research, Nagoya University \\
Furo-cho, Chikusa-ku, Nagoya, Aichi 464-8601, Japan}

\author[0000-0001-6583-2594]{Takashi Sekii}
\affiliation{National Astronomical Observatory of Japan \\
2-21-1 Osawa, Mitaka, Tokyo 181-8588, Japan}
\affiliation{Astronomical Science Program, The Graduate University for Advanced Studies, SOKENDAI\\
2-21-1 Osawa, Mitaka, Tokyo 181-8588, Japan}



\begin{abstract}
We have carried out inversions of travel times as measured by \citet{2020Sci...368.1469G} 
to infer the internal profile 
of the solar meridional circulation (MC). 
A linear inverse problem has been solved by the regularized least-squares method 
with a constraint that the angular momentum (AM) transport 
by MC should be equatorward (HK21-type constraint). 
Our motivation for using this constraint is based on the result by \citet{2021NatAs...5.1100H} 
where the solar equator-fast rotation was reproduced successfully 
without any manipulation. 
The inversion result indicates that the MC profile is a double-cell structure 
if the so-called HK21 regime, in which AM transported by MC sustains the equator-fast rotation, 
correctly describes the physics inside the solar convective zone. 
The sum of the squared residuals computed with the inferred double-cell MC profile 
is comparable to that computed with the single-cell MC profile 
obtained when we exclude the HK21-type constraint, 
showing that both profiles can explain the data more or less at the same level. 
However, we also find that adding the HK21-type constraint degrades the
resolution of the averaging kernels. 
Although it is difficult for us to determine the 
large-scale morphology of the solar MC 
at the moment, 
our attempt highlights the relevance of 
investigating the solar MC profile 
from both theoretical and observational perspectives. 



\end{abstract}



\section{Introduction} \label{sec:intro}
The meridional circulation (MC) is one 
component of the Sun's large-scale 
flows. 
On the surface of the Sun, 
the velocity amplitude of the solar MC is 
small 
\citep[$\sim 10-20 \,\mathrm{m} \, \mathrm{s}^{-1}$; e.g.][]{1996ApJ...460.1027H,2012ApJ...761L..14R}
compared with the other velocity components, 
such as granulation \citep[$\sim 10^{3} \, \mathrm{m} \, \mathrm{s}^{-1}$; e.g.][]{2009LRSP....6....2N} 
and rotation \citep[$\sim 2 \times 10^{3} \,\mathrm{m} \, \mathrm{s}^{-1}$; e.g.][]{1996Sci...272.1300T,2009LRSP....6....1H}. 
However, it nevertheless has substantial impacts on magnetic flux transport 
toward the pole \citep{1989Sci...245..712W}; 
thus, it possibly controls the periodicity of the solar activity cycle 
\citep[e.g.][]{2020LRSP...17....4C}. 
In addition, the solar MC redistributes the angular momentum 
(AM) inside the convective zone 
\citep[e.g.][]{2005LRSP....2....1M}. 
In a statistically steady state, 
the AM profile is determined by 
a balance between the amount of AM 
transported by MC and that transported by turbulent Reynolds stresses 
\citep[c.f. gyroscopic pumping,][]{2011ApJ...743...79M}. 
Therefore, we can extract information about the turbulent stresses 
in the convective zone 
once we know the internal MC profile 
because we already have quite a good picture of the AM profile in the solar convective zone as shown 
by 
global helioseismology \citep[e.g.][]{1998ApJ...505..390S}. 
As such, it is of great importance to investigate the solar internal MC profile to improve our understanding of the solar dynamo mechanism. 

One 
of the most powerful methods for inferring the solar internal MC profile 
is time--distance helioseismology 
\citep{1993Natur.362..430D,2010ARA&A..48..289G} 
in which wave travel times between 
two points on the solar surface 
may be linearly related to internal flow fields via forward modeling 
formulated based on, for example, ray approximation 
\citep[e.g.][]{1997ASSL..225..241K} 
or Born approximation \citep[e.g.][]{2002ApJ...571..966G}. 
The measured wave travel times are then inverted to 
estimate the flow fields. 

Because of the relevance of the MC in 
the solar dynamo, 
there have been multiple attempts to infer the internal MC profile 
via time--distance helioseismology 
\citep{1997Natur.390...52G,2000PhDT.........9G,2004ApJ...603..776Z,2005ApJ...630.1206C,2013ApJ...774L..29Z,2015ApJ...805..133J,2015ApJ...813..114R,2017ApJ...849..144C,2020Sci...368.1469G,2023ApJ...954..187H}. 
We have seen good agreement for the MC profile in the subsurface region 
($r/R_{\odot}>0.95$ where $r$ and $R_{\odot}$ are 
the distance from the solar center and solar radius, respectively), as 
the poleward flow persists in the subsurface region 
with a peak velocity amplitude of about $\sim 10-20$ $\mathrm{m} \, \mathrm{s}^{-1}$. 
In contrast, 
there has been no consensus for deeper regions 
\citep[see][]{2022LRSP...19....3H,2023SSRv..219...77H}. 
It is not clear whether 
the MC 
throughout the convective zone is a single- or multiple-cell structure
\citep[e.g.][]{2020Sci...368.1469G,2023ApJ...954..187H}. 
The difference should be critical for the periodicity of the solar activity cycle 
as indicated by flux-transport dynamo models 
\citep[e.g.][]{1995A&A...303L..29C,1999ApJ...518..508D}. 
It 
has been also recently pointed out by \citet{2024SciA...10K5643B} 
that a single-cell MC profile 
and tiny negative superadiabaticity around the lower convective boundary 
can produce the latitudinal entropy gradient 
which 
is consistent with observed properties of the 
solar inertial modes \citep{2021A&A...652L...6G}. 

A primary reason for the difficulty in 
estimating the internal MC profile in the deep convective zone is that 
the MC velocity amplitude is 
expected to be so small there 
\citep[$\sim$ a few $\mathrm{m} \, \mathrm{s}^{-1}$;][]{2000PhDT.........9G} 
that the travel time perturbations 
caused by MC 
($\sim 0.1 \, \mathrm{s}$) 
are small and comparable to the observational uncertainties 
that are dominated by realization noises 
\citep{2008ApJ...689L.161B,2021ApJ...911...90S}. 
Inversion results are thus quite sensitive to the noise in the travel time perturbations. 
Although we can somehow obtain solutions 
by imposing constraints or regularization 
in inversion procedures, 
there is no 
unique method for imposing constraints 
(or how we add prior information), 
resulting in 
a number of possible solutions. 

Given the current level of the low signal-to-noise ratio 
\citep[that can be possibly increased by some future solar polar missions;][]{2023BAAS...55c.105F} 
as well as the difficulty in MC inversion, 
improving on the inversion procedures is certainly valuable. 
For example, 
\citet{2023ApJ...954..187H} 
applied Bayesian statistics to carry out MC inversion, 
which has enabled them to partly avoid the 
arbitrary choice of some regularization terms. 
Such attempts to reconsider 
inversion procedures 
from various perspectives will be helpful 
for analyzing the data 
introduced by future observations. 

In this paper, we focus on the inversion procedure. 
Specifically, we carry out MC inversion 
by imposing 
constraints 
inspired by the results from numerical simulations of solar convection 
in addition to those adopted conventionally, such as the mass conservation constraint. 

There are currently two 
main results widely accepted 
that successfully reproduce the solar differential rotation where 
the equator rotates faster than the pole (i.e., equator-fast rotation). 
In one regime, 
rotation has significant impacts on turbulent convection, 
and the resultant AM transport by Reynolds stress is crucial 
for achieving
equator-fast rotation 
\citep[e.g.][]{2014MNRAS.438L..76G,2015ApJ...804...67F,2015A&A...576A..26K,2023A&A...669A..98K}, 
called the RS regime in this work. 
However, it should be noted that 
we cannot realize the RS regime 
without adopting 
strong effective diffusion and viscosity 
that do not seem to be compatible with realistic solar parameters. 

Another regime is the so-called HK21 regime that 
has been proposed by \citet{2021NatAs...5.1100H} 
who reproduce the solar equator-fast rotation without any manipulation. 
In the HK21 regime, 
the magnetic field plays an important role in AM redistribution, 
and the equator-fast rotation is sustained by the equatorward AM transport by MC 
\citep{2021NatAs...5.1100H,2022ApJ...933..199H}. 
Because the AM transport by MC 
is directly related to the meridional flow field, 
the suggested equatorward AM transport by MC 
can be used as 
a physically motivated constraint for the MC profile 
if the physics found in the HK21 regime is assumed to be correct. 
Therefore, as a first step toward implementing physics constraints on MC inversion, 
we would like 
to infer the large-scale morphology 
of the solar MC profile 
with the constraint on AM transport by MC. 

This paper is structured as follows. 
In Section \ref{sec:data}, we explain 
the data used in MC inversion. 
We then present how we invert the travel times to infer the MC profile 
(Section \ref{sec:method}) 
where 
we introduce constraints 
devised based on the numerical results found in the HK21 regime 
(hereafter the HK21-type constraint). 
%
The results for MC inversion 
that are obtained with or without the HK21-type constraint 
are shown in Section \ref{sec:result}. 
Although the main focus of this paper is 
on inversion with a constraint on AM transport by MC, 
we also carry out inversion with a constraint 
appropriate for the RS regime (hereafter the RS-type constraint) 
in a rather crude manner. 
We discuss all the inversion results obtained in this study 
in Section \ref{sec:disc}. 
We conclude in Section \ref{sec:conc}. 

\section{Data and kernels} \label{sec:data}
To carry out MC inversion, 
we used the travel times, variance--covariance matrix, 
and sensitivity kernels, 
which are all publicly available from \citet{2020Sci...368.1469G,3.NAPBUA_2020} 
\dataset[(Open Research Data Repository of the Max Planck Society)]
{https://edmond.mpdl.mpg.de/imeji/collection/0MJjNql7GfpEl5Mb}. 
We give a brief summary of the data in the following few paragraphs 
\citep[see][hereafter G20]{2020Sci...368.1469G}. 

We first explain how the travel times are computed in G20. 
In general, the starting point for measuring travel times is an
observation of the surface velocity fields 
of the Sun. 
G20 used two series of Dopplergrams, namely, 
the one obtained by the Solar and Heliospheric Observatory (SOHO)/Michelson Doppler Imager (MDI) 
\citep{1995SoPh..162..129S} 
for cycle 23 
(duration $\sim 11 \, \mathrm{years}$)
and the other obtained by the Global Oscillation Network Group (GONG) 
\citep{1996Sci...272.1292H} 
for cycle 24 
(duration $\sim 11 \, \mathrm{years}$). 

After substantial data reduction efforts 
to render the raw Dopplergrams 
suitably for time--distance helioseismic analyses 
\citep[see][]{2017A&A...601A..46L,2018A&A...619A..99L}, 
cross-correlation functions of the velocity fields 
for multiple pairs of points on the solar surface 
are computed. 
These point pairs aligned in the north--south direction were chosen 
to measure travel times that are sensitive to MC. 
Spatial averaging of the cross-correlation functions 
has been conducted to reduce the realization noise 
in travel time measurements. 
This spatial averaging has been characterized by 
two parameters, namely,  
$\lambda$, the latitude of the center 
of the circle whose northern and southern arcs are used 
for spatial averaging, and $\Delta$, 
the angular distance between the two opposing arcs. 
The details for the data reduction, arc geometry, 
and how the spatial averaging has been done 
can be found in \citet{2017A&A...601A..46L,2018A&A...619A..99L}. 

Then, with the cross-correlation functions 
averaged spatially, 
G20 applied a one-parameter fitting method 
to compute the north--south travel times 
\citep{2002ApJ...571..966G,2004ApJ...614..472G}. 
The computed travel times are given as a function of 
$\lambda$ and $\Delta$ 
that were introduced 
in the previous paragraph. 
The parameter ranges are 
$-54^{\circ} < \lambda < 54^{\circ}$, where the positive (negative) value represents the 
northern (southern) hemisphere, 
and 
$6^{\circ} < \Delta < 42^{\circ}$ (in units of degrees in both cases). 
The total number of travel times thus computed is 
$N_{d} = 9120$ for each of the MDI and GONG data. 
The variance--covariance of the travel times is also evaluated by 
G20 
\citep[see also][]{2004ApJ...614..472G,2014A&A...567A.137F}. 

We then introduce a series of equations 
to infer the internal MC profile with the measured travel times. 
First, the forward problem must be solved 
\citep{2002ApJ...571..966G}. 
G20 relied on the Born approximation, which could be used to relate 
the $i$-th component of the north--south travel times 
$\tau_i$ to 
the meridional flow field as: 
\begin{eqnarray}
\tau_{i} & = & \iint (\mathcal{K}_{i}^{r}(r,\theta) U_{r}(r,\theta) + \mathcal{K}_{i}^{\theta}(r,\theta) U_{\theta}(r,\theta)) \mathrm{d}r \mathrm{d}\theta \label{eq_td_intg} \\ 
(i & = & 1, ..., N_{d}; N_{d} = 9120), \nonumber 
\end{eqnarray}
where 
$\theta$ is the 
colatitude. 
The radial and latitudinal components of the meridional flow field are 
denoted by $U_{r}$ and $U_{\theta}$. 
The corresponding sensitivity kernels are 
$\mathcal{K}_{i}^{r}$ ($\mathcal{K}_{i}^{\theta}$) 
that have been calculated 
by numerically solving wave propagation in 
a particular solar model \citep{2017A&A...600A..35G}. 
Axisymmetry is assumed to derive the presented equation 
\citep{2018A&A...616A.156F}. 
Index $i$ designates a certain pair $(\lambda, \Delta)$, 
namely, $i$ runs from $1$ to $N_{d}$ in the case of G20's data. 
The integration is conducted over the whole solar interior 
(although the meridional flow velocity is assumed to be zero 
outside the convective zone, as mentioned in Section \ref{sec:RLS}.) 

To parameterize the meridional flow field, 
G20 expanded it 
in both the radial and latitudinal directions 
with the cubic B-spline functions $Q_{k}(r)$ ($k=1,...,N_{k}$; $N_{k}=20$)
and Legendre polynomials $P_{l}(\mathrm{cos} \, \theta)$ ($l=0,...,N_{l}-1$; $N_{l} = 16$), respectively. 
The $18$ knots used for defining the B-spline function 
are uniformly spaced between the bottom of the convective zone 
and the solar surface. 
The explicit expression of the expanded meridional flow field is then 
\begin{eqnarray}
U_{s}(r,\theta) & = & \sum_{j=(k,l)} u_{s,j} Q_{k}(r) P_{l}(\mathrm{cos} \, \theta) \label{eq_MCv_expanded} \\
(s & = & r \, \, \mathrm{or} \, \, \theta) \nonumber \\  
(k & = & 1, ..., N_{k}; N_{k} = 20) \nonumber \\ 
(l & = & 0, ..., N_{l}-1; N_{l} = 16), \nonumber 
\end{eqnarray}
in which index $s$ represents 
$r$ or $\theta$. 
A certain pair $(k,l)$ is represented by the index $j$ 
that runs from $1$ to $N_{j}$ 
where $N_{j} = N_{k} \times N_{l}$. 
The expansion coefficient is denoted by $u_{s,j}$. 
By defining an $N_{j}$-dimensional vector $\boldsymbol{u}_{r}$ 
($\boldsymbol{u}_{\theta}$) 
whose $j$-th element is $u_{r,j}$ ($u_{\theta,j}$), 
Equation (\ref{eq_td_intg}) can be rewritten in matrix form: 
\begin{equation}
\boldsymbol{\tau} = K \boldsymbol{u} + \boldsymbol{e}, \label{eq_td_mtx}
\end{equation}
where $\boldsymbol{u}$ is a $2N_{j}$-dimensional vector defined as: 
\begin{eqnarray}
\boldsymbol{u} 
= 
\begin{pmatrix}
\boldsymbol{u}_{r} \\
\boldsymbol{u}_{\theta} 
\end{pmatrix}. \nonumber \\ 
 \label{eq_vecu} 
\end{eqnarray} 
The $N_{d}$-dimensional vector $\boldsymbol{\tau}$ represents 
the measured travel times. 
Note that the $N_{d}$-dimensional vector $\boldsymbol{e}$ is added in the expression 
as an observational error. 
The statistical property of $\boldsymbol{e}$ 
is described by the $N_{d} \times N_{d}$ variance--covariance matrix ($E$) of the travel time. 
In Equation (\ref{eq_td_mtx}), the sensitivity kernel is represented by the $N_{d} \times 2N_{j}$ matrix $K$ 
consisting of two $N_{d} \times N_{j}$ matrices $K^{r}$ and $K^{\theta}$ 
that are defined as 
\begin{equation}
(K^{s})_{ij} = \iint \mathcal{K}^{s}_{i}(r,\theta) Q_{k}(r) P_{l}(\mathrm{cos} \, \theta) \mathrm{d}r \mathrm{d}\theta, \label{eq_krn_rth_expanded}
\end{equation}
with which $K = (K^{r} \, K^{\theta})$. 
The index $s$ again denotes $r$ or $\theta$. 
Note that $j = (k,l)$. 

The measured travel times, variance--covariance matrix, 
basis functions, and expanded kernels 
are provided by G20. 
Using the data described in this section, 
the matrix equation (\ref{eq_td_mtx}) is to be inverted 
with some regularization and constraints on the MC profile, 
as shown in Section \ref{sec:method}. 

\section{Method} \label{sec:method}
In this section, we explain how to invert 
Equation (\ref{eq_td_mtx}). 
Basically, 
the regularized least-squares (RLS) method is used, 
thus, we first briefly review the RLS method 
in Section \ref{sec:RLS}. 
We then present 
a constraint on MC inversion that is devised 
based on the result of \citet{2021NatAs...5.1100H} 
(hereafter HK21) 
(see Section \ref{sec:intro}) 
in Section \ref{sec:RLS_HK}. 
Because we have free parameters in the inversion procedure, 
which are called trade-off parameters, 
the process for determining these parameters is given in Section \ref{sec:how2_reason}. 

\subsection{The regularized least-squares method} \label{sec:RLS}
It is often the case that 
the least-squares solution of Equation (\ref{eq_td_mtx}) 
is unstable against observational uncertainties and/or 
numerical 
errors; therefore, 
the inverse problem is ill-posed. 
However, an ill-posed problem can be turned into 
a well-posed one 
by adding prior information 
such as that the solution should be flat or smooth 
enough to exhibit no 
discontinuous features which are difficult 
to accept from the physics point of view. 
This procedure is called regularization. 

In the RLS method \citep{TA77}, 
what we minimize is: 
\begin{equation}
X = | E^{-1/2} (\boldsymbol{\tau} - K \boldsymbol{u}) |^{2} 
+ 
\alpha |D \boldsymbol{u}|^2 
, \label{eq_S_RLS}
\end{equation}
where the first term on the right-hand side represents the sum of squared residuals 
between the observation ($\boldsymbol{\tau}$) and model ($K \boldsymbol{u}$) 
normalized by the variance--covariance matrix ($E$). 
The second term represents the regularization term. 
For example, if the matrix $D$ is a first- (second-)derivative operator, 
the term $D \boldsymbol{u}$ represents 
the flatness (smoothness) of the flow field 
\citep[e.g.][]{2015ApJ...813..114R}. 
In this study, we follow G20 and 
define an $N_{D} \times 2N_{j}$ matrix $D$ 
(where $N_{D}$ is the number of mesh points 
in the $(r,\theta)$-plane) 
so that $D \boldsymbol{u}$ represents 
the weighted vorticity of the MC flow field 
(see the supplementary material of G20, 
or Appendix \ref{sec:app:c} of this paper for more details). 
The balance between the first and second terms in Equation (\ref{eq_S_RLS}) 
depends on the trade-off parameter $\alpha$. 
We obtain the least-squares solution in the limit $\alpha \rightarrow 0$. 
Conversely, the solution is determined by the prior 
information alone 
when $\alpha \rightarrow \infty$. 
For a particular value of $\alpha$, 
we can obtain the corresponding RLS solution 
that minimizes the quantity $X$. 
Note that 
there is no established way of choosing 
a single most appropriate value of $\alpha$. 
We discuss the point in Section \ref{sec:how2_reason}. 

Besides regularization, we consider two additional constraints on the meridional flow field 
that are conventionally adopted in some previous studies 
\citep[e.g.][]{2015ApJ...813..114R,2020Sci...368.1469G,2023ApJ...954..187H}. 
One is the mass conservation $\nabla \cdot (\rho \boldsymbol{U}) = 0$ 
where $\boldsymbol{U} = (U_{r}\, U_{\theta})^{\mathrm{T}}$, 
and the other one is given based on the assumption that 
MC is confined in the solar convective zone and 
does not cross the convective boundary. 
Both constraints can be expressed as linear constraints 
(in terms of the expansion coefficients $\boldsymbol{u}$) as follows: 
%
\begin{equation}
C \boldsymbol{u} = \boldsymbol{0} \label{eq_constrain_mass}
\end{equation}
for the former, and
\begin{equation}
S \boldsymbol{u} = \boldsymbol{0} \label{eq_constrain_conv}
\end{equation} 
for the latter. 
The detailed derivation of the matrices $C$ and $S$ is found in 
the supplementary material in G20. 
We have used $C$ and $S$ that are provided by G20. 
It should be noted that these two constraints are strict 
in a sense that Equations 
(\ref{eq_constrain_mass}) and (\ref{eq_constrain_conv}) should 
always be satisfied. 

Then, what we must minimize with the strict constraints is: 
\begin{equation}
X' = | E^{-1/2} (\boldsymbol{\tau} - K \boldsymbol{u}) |^{2} 
+ 
\alpha |D \boldsymbol{u}|^2  
+ 
\boldsymbol{\kappa} \cdot C \boldsymbol{u} 
+ 
\boldsymbol{\mu} \cdot S \boldsymbol{u} 
, \label{eq_S_RLS2}
\end{equation}
where we have two additional terms 
(see Equation (\ref{eq_S_RLS})) 
with Lagrange multipliers $\boldsymbol{\kappa}$ and $\boldsymbol{\mu}$. 
Minimization of the quantity $X'$ 
is achieved by the vectors 
$\boldsymbol{u}$, $\boldsymbol{\kappa}$, and $\boldsymbol{\mu}$ 
that satisfy the following matrix equation: 
\begin{eqnarray}
\begin{pmatrix}
A^{\,} & C^{\mathrm{T}} & S^{\mathrm{T}} \\
C & O & O \\
S & O & O
\end{pmatrix}
\begin{pmatrix}
\boldsymbol{u} \\
\boldsymbol{\kappa} \\
\boldsymbol{\mu}
\end{pmatrix}
= 
\begin{pmatrix}
K^{\mathrm{T}} E^{-1} \boldsymbol{\tau} \\
\boldsymbol{0} \\
\boldsymbol{0} 
\end{pmatrix}, \nonumber \\ 
 \label{eq_mtx_eq} 
\end{eqnarray} 
where we have introduced a $2 N_{j} \times 2 N_{j}$ submatrix $A$ 
that is defined as: 
\begin{eqnarray}
A = K^{\mathrm{T}}E^{-1}K + \alpha D^{\mathrm{T}}D. \label{eq_submtx_A} 
\end{eqnarray} 
This equation is derived by equating the gradient of $X'$ to $\boldsymbol{0}$. 
Let us denote by $\hat{\boldsymbol{u}}_{\mathrm{RLS}}$ 
the RLS solution of Equation (\ref{eq_td_mtx}) 
obtained with the constraints 
(\ref{eq_constrain_mass}) and (\ref{eq_constrain_conv}). 

Before concluding this section, 
we would like to mention that 
without any other constraints, 
we have obtained the RLS solution $\hat{\boldsymbol{u}}_{\mathrm{RLS}}$ 
that exhibits the single-cell MC profile; hence, 
we have confirmed the same result as that found by G20. 
We discuss whether this result changes or not 
later in Section \ref{sec:result} 
when we further add the HK21-type constraint. 

\subsection{Constraint on the MC profile devised based on numerical results found in the HK21 regime \\ 
(the HK21-type constraint)} \label{sec:RLS_HK}
\begin{figure}[t]
\begin{center}
\includegraphics[scale=0.5]{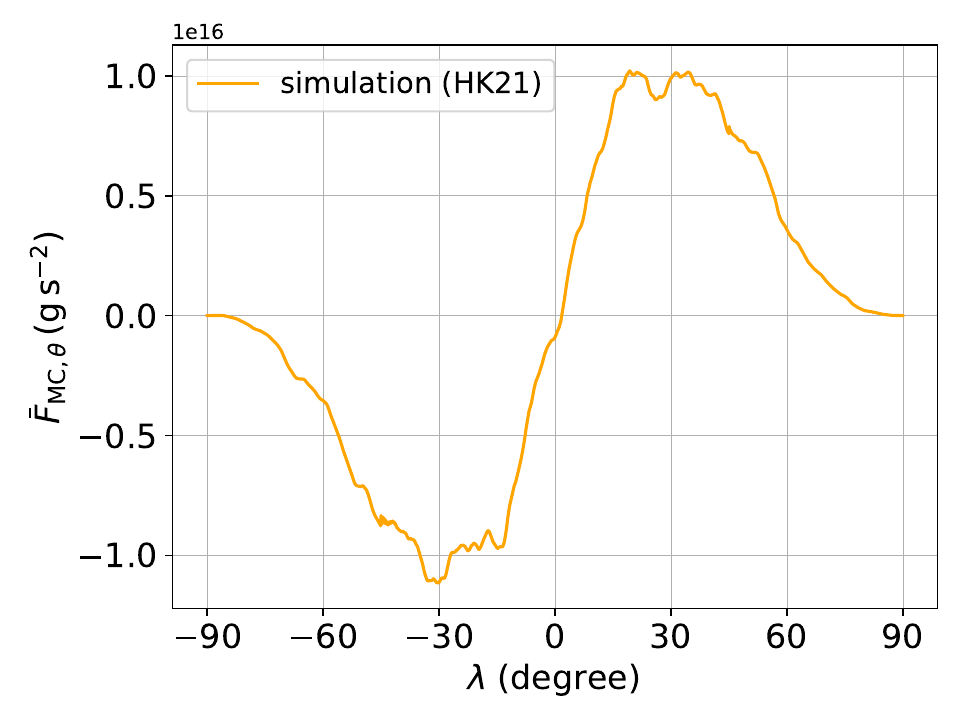}
\caption{\footnotesize Latitudinal AM flux by MC that is radially averaged 
$\bar{F}_{\mathrm{MC},\theta}(\theta)$ 
as a function of the latitude ($\lambda = \pi / 2 - \theta$ 
where $\theta$ is the colatitude). 
The MC profile with the highest-resolution simulation in HK21 
has been used to compute the AM flux 
(see Equation (\ref{eq_A3_Fmc_th})). 
Note that the positive $\bar{F}_{\mathrm{MC},\theta}(\theta)$ 
indicates that 
the AM flows in the direction with the larger colatitude; that is, 
the AM is transported southward (northward) in the northern (southern) hemisphere 
if we take $\theta=0$ as the northern pole. 
}
\label{fig:m1}
\end{center}
\end{figure}

In this section, we present how to impose the HK21-type constraint 
on Equation (\ref{eq_td_mtx}). 
As mentioned in Section \ref{sec:intro}, 
the magnetic field plays a significant role in AM transport in this regime, 
and the equator-fast rotation is sustained by the AM transported by the MC flow 
\citep{2022ApJ...933..199H}. 
This can be confirmed by looking at 
the latitudinal AM flux by MC 
that is defined as: 
\begin{equation}
F_{\mathrm{MC},\theta} = \rho \mathcal{L} U_{\theta}, \label{eq_AMF_MC}
\end{equation} 
in which the density and specific AM are 
denoted by $\rho$ and $\mathcal{L}$, respectively. 
Note that $\mathcal{L} = r \, \mathrm{sin} \, \theta \, v_{\phi}$ 
where $v_{\phi}$ represents the rotational velocity 
(as a function of $r$ and $\theta$) in the inertial frame. 
Figure \ref{fig:m1} shows the latitudinal AM flux by MC 
radially averaged 
(hereafter $\bar{F}_{\mathrm{MC},\theta}(\theta)$) 
that has been computed with the result of numerical simulation by HK21, 
highlighting the fact that the MC transports the AM toward the equatorial region 
in the HK21 regime. 
See Equation (\ref{eq_A3_Fmc_th}) 
for the definition of $\bar{F}_{\mathrm{MC},\theta}(\theta)$.) 

Because the latitudinal AM flux ($F_{\mathrm{MC},\theta}$) 
is proportional to the latitudinal component of the meridional flow field ($U_{\theta}$), 
we can readily obtain a linear constraint for the meridional flow field. 
In this study, we first assume that 
the latitudinal average of 
$\bar{F}_{\mathrm{MC},\theta}(\theta)$ 
is positive (negative) 
in the northern (southern) hemisphere 
to reproduce the net AM transport by MC 
toward the equator (Figure \ref{fig:m1}). 
The assumption 
can be expressed as: 
\begin{equation}
D_{\mathrm{HK1}} \boldsymbol{u} = \boldsymbol{b}, \label{eq_constrain_AM}
\end{equation} 
where $D_{\mathrm{HK1}}$ is a $2 \times 2 N_{j}$ matrix 
with which $D_{\mathrm{HK1}} \boldsymbol{u}$ represents 
latitudinal averages of $\bar{F}_{\mathrm{MC},\theta}$ 
in the northern and southern hemispheres. 
For the vector $\boldsymbol{b}$, 
we have used the HK21 results, namely, 
$\boldsymbol{b} = (5 \times 10^{15}, -5 \times 10^{15})$ 
in units of $\mathrm{g} \, \mathrm{s}^{-2}$. 
Note that we do not find any significant difference in the
large-scale morphologies of inferred MC profiles 
even when we change the values of the elements in the vector $\boldsymbol{b}$ 
by a few orders, 
as discussed in Section \ref{sec:disc1}. 

Second, to ensure the equatorward AM fluxes at almost all the latitudes, 
we further assume that $\bar{F}_{\mathrm{MC},\theta}(\theta)$ 
is not rapidly varying in the latitudinal direction 
(see Figure \ref{fig:m1}). 
We thus consider a regularization term 
$D_{\mathrm{HK2}} \boldsymbol{u}$ 
that represents the latitudinal derivative of 
$\bar{F}_{\mathrm{MC},\theta}(\theta)$. 
The regularization matrix $D_{\mathrm{HK2}}$ is an $N_{\theta} \times 2 N_{j}$ matrix 
where $N_{\theta}$ is the number of grid points in the latitudinal direction. 
We present how to compute the matrices $D_{\mathrm{HK1}}$ 
and $D_{\mathrm{HK2}}$ in 
Appendix \ref{sec:app:c}. 


When considering the new constraints on AM transport by MC, 
the quantity to be minimized is: 
\begin{eqnarray}
X_{\mathrm{HK}}' & = & | E^{-1/2} (\boldsymbol{\tau} - K \boldsymbol{u}) |^{2} 
+ 
\alpha |D \boldsymbol{u}|^2  
+ 
\boldsymbol{\kappa} \cdot C \boldsymbol{u} 
+ 
\boldsymbol{\mu} \cdot S \boldsymbol{u} \nonumber \\ 
& \, & + 
\beta |D_{\mathrm{HK1}} \boldsymbol{u} - \boldsymbol{b}|^{2} 
+ 
\gamma |D_{\mathrm{HK2}} \boldsymbol{u}|^{2} 
, \label{eq_S_RLS_AM}
\end{eqnarray}
where new trade-off parameters $\beta$ and $\gamma$ have been introduced. 
Note that the new constraints are taken as regularization terms 
rather than strict constraints; hence, 
we must check whether a solution obtained with a certain set of trade-off parameters 
satisfies the constraints, as explained in Section \ref{sec:how2_reason}. 
The solution for the minimization of $X_{\mathrm{HK}}'$ 
can be obtained by solving the following matrix equation 
(in almost the same way as that when we have derived Equation (\ref{eq_mtx_eq})): 
\begin{equation}
\begin{pmatrix}
A' & C^{\mathrm{T}} & S^{\mathrm{T}} \\
C & O & O \\
S & O & O
\end{pmatrix}
\begin{pmatrix}
\boldsymbol{u} \\
\boldsymbol{\kappa} \\
\boldsymbol{\mu}
\end{pmatrix}
= 
\begin{pmatrix}
K^{\mathrm{T}} E^{-1} \boldsymbol{\tau} + \beta D_{\mathrm{HK1}}^{\mathrm{T}} \boldsymbol{b} \\
\boldsymbol{0} \\
\boldsymbol{0} 
\end{pmatrix}, \label{eq_mtx_eq_AM} 
\end{equation}
where we have introduced a $2 N_{j} \times 2 N_{j}$ submatrix $A'$ 
that is defined as: 
\begin{eqnarray}
A' = A + \beta D_{\mathrm{HK1}}^{\mathrm{T}}D_{\mathrm{HK1}} + \gamma D_{\mathrm{HK2}}^{\mathrm{T}}D_{\mathrm{HK2}} \label{eq_submtx_Ap} 
\end{eqnarray} 
(see Equation (\ref{eq_submtx_A}) for the definition of $A$). 
It should be noted that 
in addition to the $2N_{j}$-dimensional data vector $K^{\mathrm{T}} E^{-1} \boldsymbol{\tau}$, 
we have an extra $2N_{j}$-dimensional vector 
$\beta D_{\mathrm{HK}}^{\mathrm{T}} \boldsymbol{b}$ 
on the right-hand side of Equation (\ref{eq_mtx_eq_AM}) 
that originates from the regularization term newly introduced. 
The solution thus obtained is shown later in Section \ref{sec:result}. 

\subsection{How to determine values of the trade-off parameters 
and choose reasonable solutions} \label{sec:how2_reason}
As described in Sections \ref{sec:RLS} and \ref{sec:RLS_HK}, 
once we determine values of the trade-off parameters, 
the RLS solution for Equation (\ref{eq_td_mtx}) 
with the HK21-type constraint 
can be obtained by solving the matrix equation 
(\ref{eq_mtx_eq_AM}). 
Here, we present how we determined the values 
of the trade-off parameters 
among numerous possible parameters. 
We will hereafter call the solutions thus obtained 
``reasonable solutions.'' 


First, we prepare a grid of the trade-off parameters. 
In the case of the HK21 regime, 
the decimal logarithms of $\alpha$, 
$\beta$, and $\gamma$ range from $-3.5$ to $-2$, 
from $-34$ to $-26$, 
and from $-36$ to $-27$, respectively. 
Note the units of the trade-off parameters; that is,
we express 
$r$, $\rho$, and $\boldsymbol{U}$ 
in units of $\mathrm{m}$, $\mathrm{g} \, \mathrm{m}^{-3}$, 
and $\mathrm{m} \, \mathrm{s}^{-1}$, respectively. 
The range of $\alpha$ is determined following the G20 results, and 
those of $\beta$ and $\gamma$ are chosen so that the inverse problem (\ref{eq_td_mtx}) 
is well-posed. 
For each set of the trade-off parameters  
in the prepared grid, 
we obtain the corresponding solution by solving Equation 
(\ref{eq_mtx_eq_AM}). 
We denote the solution thus obtained by $\hat{\boldsymbol{u}}_{\mathrm{cand}}$. 

Among the candidate solutions $\hat{\boldsymbol{u}}_{\mathrm{cand}}$, 
we have chosen reasonable solutions 
based on the following two criteria. 
The first criterion is whether a candidate solution satisfies the constraint 
we have assigned in Section $\ref{sec:RLS_HK}$. 
We need this criterion because what we have imposed 
on the inverse problem (Equation (\ref{eq_td_mtx})) 
is not a strict constraint but regularization 
(see Section \ref{sec:RLS_HK}). 
Specifically, when we consider the HK21-type constraint, 
we exclude candidate solutions that exhibit poleward $\bar{F}_{\mathrm{MC},\theta}(\theta)$ 
in either hemisphere. 

The second criterion is 
whether we can distinguish the large-scale morphology 
of an inferred MC profile, namely, a single- or multiple-cell structure, with 2-$\sigma$ significance. 
Estimated uncertainties are evaluated by introducing 
a $2 N_{j} \times N_{d}$ matrix $R_{\mathrm{cand}}$ 
with which $\hat{\boldsymbol{u}}_{\mathrm{cand}} = R_{\mathrm{cand}} \boldsymbol{\tau}$. 
The uncertainty of the $i$-th component in $\hat{\boldsymbol{u}}_{\mathrm{cand}}$ is then: 
\begin{equation}
\delta (\hat{\boldsymbol{u}}_{\mathrm{cand}})_{i} = \biggl ( \sum_{j,k} (R_{\mathrm{cand}})_{ij} E_{jk} (R_{\mathrm{cand}})_{ik} \biggr )^{\frac{1}{2}}, \label{eq_est_err}
\end{equation} 
based on which estimated uncertainties for the meridional flow field 
(that is computed by using $\hat{\boldsymbol{u}}_{\mathrm{cand}}$ as expansion coefficients, 
see Equation (\ref{eq_MCv_expanded})) 
can be calculated. 

In the next section, we show some of the reasonable solutions 
and the corresponding sets of trade-off parameters. 
It should be noted that we have not determined 
the most appropriate solution among the reasonable ones, 
which is in principle possible by, for example, 
selecting a solution that realizes a minimum 
residual term 
(see the first term of the right-hand side in, e.g., Equation (\ref{eq_S_RLS})). 
The reason why we are satisfied with just the reasonable solutions 
is that 
the goal of this paper is 
to infer the large-scale morphology of the MC profile 
rather than to determine detailed structures of the meridional flow field. 
It is worth mentioning that 
the results show slight differences in the MC profiles of 
the reasonable solutions, 
highlighting little relevance of an attempt to find the ``most appropriate'' solution in this study. 
\begin{figure}[t]
\begin{center}
\includegraphics[scale=0.5]{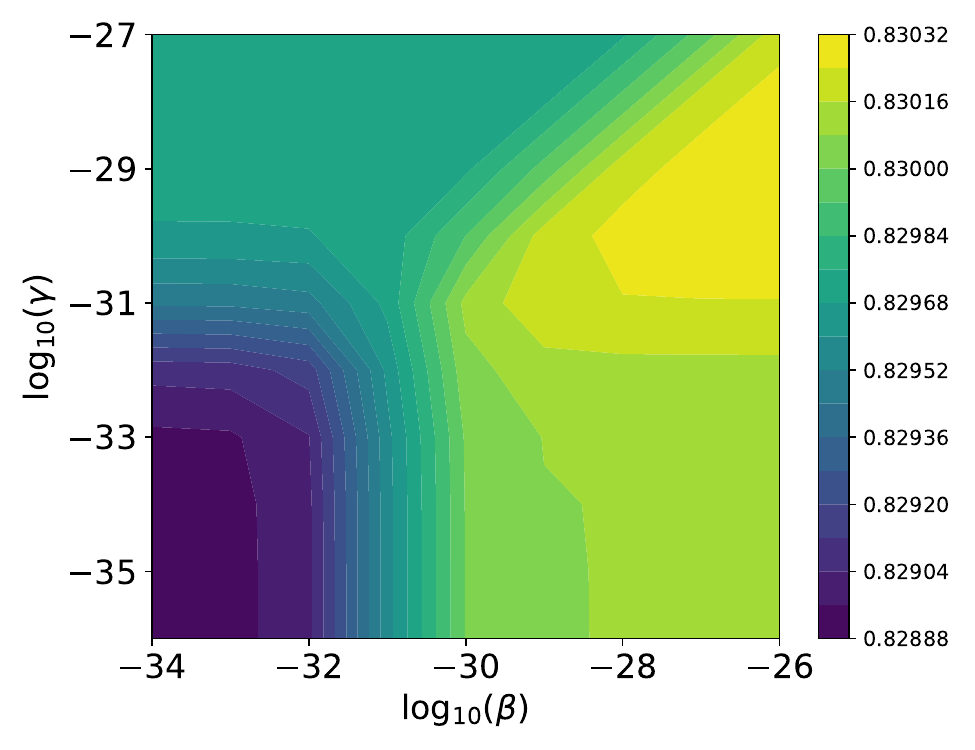}
\caption{\footnotesize 
Trade-off plane obtained via MC inversion 
with the HK21-type constraint, 
which shows the residual term $|E^{-1/2}(\boldsymbol{\tau} - K \hat{\boldsymbol{u}}_{\mathrm{cand}})|^2/N_d$ 
as a function of the decimal logarithms of 
the trade-off parameters $\beta$ (horizontal axis) and $\gamma$ (vertical axis) 
for $\alpha = 10^{-3}$. 
The brighter colors correspond to larger values of the residual term. 
See the main text for the meanings of the variables. 
}
\label{fig:0}
\end{center}
\end{figure}

\section{Results} \label{sec:result}
With the data and method presented 
in Sections \ref{sec:data} and \ref{sec:method}, 
we have carried out inversion to infer the solar MC profile. 
In this section, we show inversion results thus obtained. 
We firstly show a trade-off relation 
to see an extent to which residuals vary with trade-off parameters (Section \ref{sec:trd_pln}). 
Secondly, we present examples of reasonable solutions 
obtained with or without the HK21-type constraint (Section \ref{sec:results}). 
We then compare the inferred MC profiles in Section \ref{sec:comparison}. 
The inversion results are available 
at \dataset[https://doi.org/10.5281/zenodo.10893108]{https://doi.org/10.5281/zenodo.10893108} 
\citep{hatta_2024_10893108}. 

\begin{figure}[t]
\begin{center}
\includegraphics[scale=0.62]{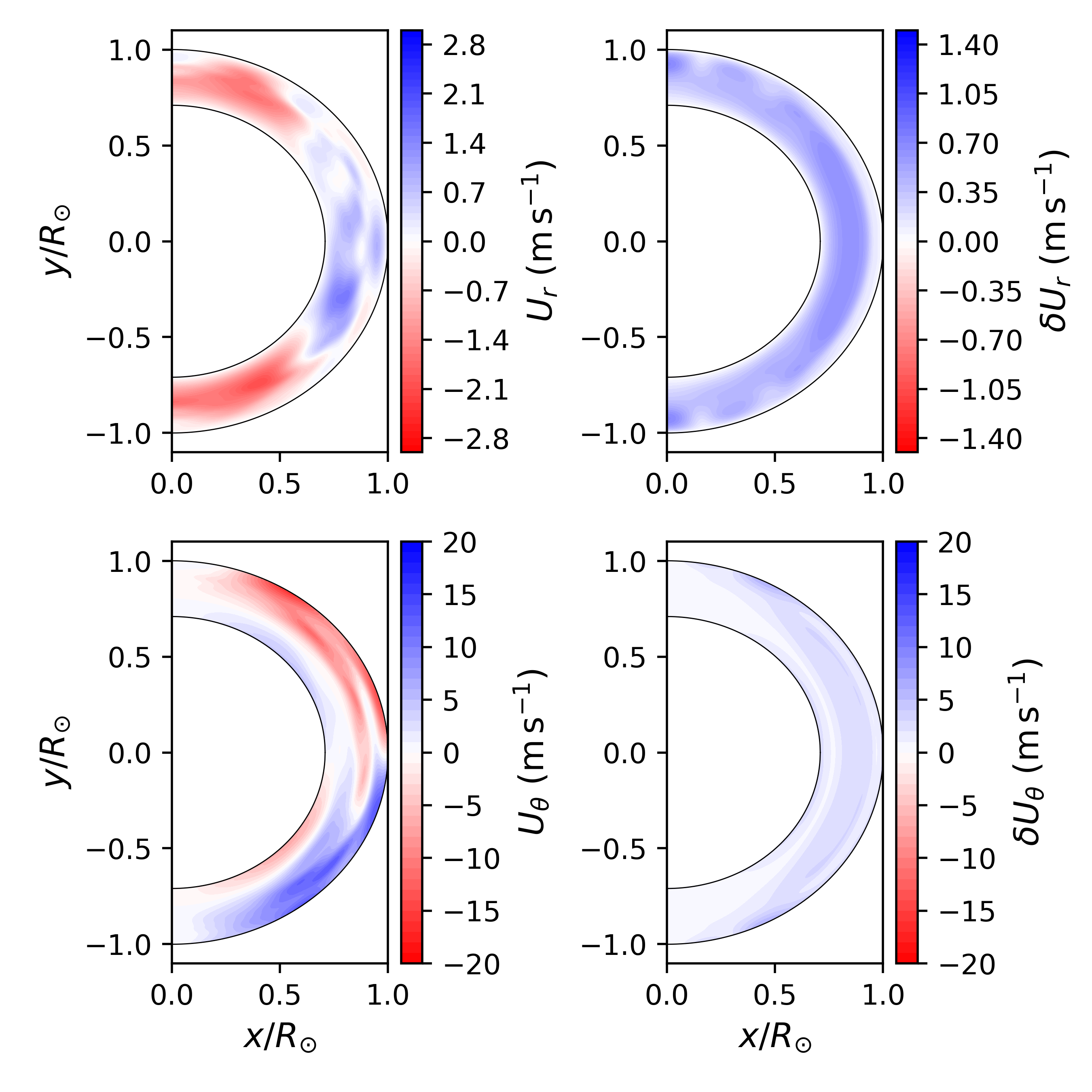}
\caption{\footnotesize Meridional flow fields 
in units of $\mathrm{m} \, \mathrm{s}^{-1}$ (left panels) 
and the estimated uncertainties 
that are determined by Equation (\ref{eq_est_err}) (right panels) 
in a meridional plane. 
The results are obtained via MC inversion without the HK21-type constraint, 
that corresponds to G20 results. 
The trade-off parameter $\alpha = 10^{-3}$, 
$\beta = 0$, and $\gamma = 0$ are 
used. 
The 2-dimensional Cartesian coordinate $(x,y)$ is 
used to express the meridional plane 
where the convective boundaries are indicated by the black curves. 
The upper-left panel shows the radial component of the inferred meridional flow field $U_{r}$ 
where red and blue indicate radially inward and outward motions, respectively. 
The lower-left panel shows the latitudinal component of the inferred meridional flow field $U_{\theta}$ 
in which red and blue indicate poleward and equatorward motions, respectively. 
A single-cell MC profile has been inferred, as has already been found by G20 
(see, e.g., Figure 2\textit{a} in G20). 
Note that we have adopted the assumption that 
MC is confined in the convective zone; 
thus, the velocity is zero elsewhere in the radiative zone below the convective zone ($r/R_{\odot} < 0.71$). 
}
\label{fig:3_rev}
\end{center}
\end{figure}

\subsection{Trade-off relation} \label{sec:trd_pln}
A trade-off relation 
between the minimization of the residual term 
(i.e., the first term in the right-hand side of Equation (\ref{eq_S_RLS_AM})) 
and regularization 
(as represented by the second, fifth, and sixth terms in the right-hand side of Equation (\ref{eq_S_RLS_AM})) 
is useful for checking 
behaviors of candidate solutions $\hat{\boldsymbol{u}}_{\mathrm{cand}}$ 
obtained for the prepared grid of the trade-off parameters 
(see Section \ref{sec:how2_reason}). 
Figure \ref{fig:0} shows the contour map of the residual term 
divided by the number of the data 
($|E^{-1/2}(\boldsymbol{\tau} - K \hat{\boldsymbol{u}}_{\mathrm{cand}})|^2/N_{d}$) 
as a function of the trade-off parameters $\beta$ and $\gamma$ 
for 
$\alpha = 10^{-3}$. 
Basically, 
the residual term increases as $\beta$ or $\gamma$ increases. 
This results from the fact that 
as we increase the trade-off parameters, 
minimizing the corresponding regularization terms is more prioritized 
over minimizing the residual term, as we mentioned in Section \ref{sec:RLS} 
(see also Equation (\ref{eq_S_RLS_AM})). 
We find the same $(\beta,\gamma)$ dependence of the residual term 
in the case of other values of $\alpha$, 
in which the absolute value of the residual becomes smaller (larger) 
when $\alpha$ becomes smaller (larger). 

\begin{figure}[t]
\begin{center}
\includegraphics[scale=0.72]{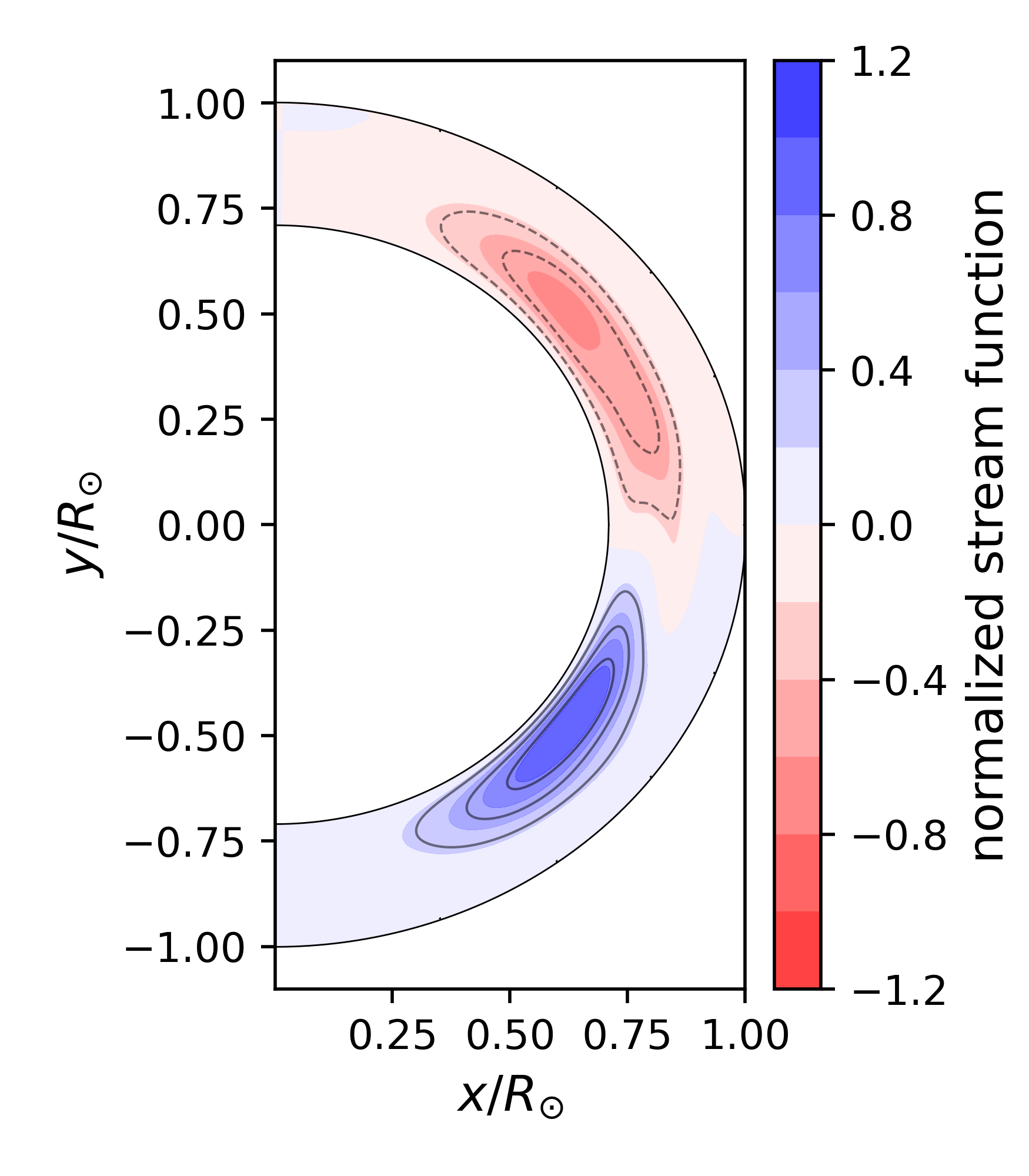}
\caption{\footnotesize Normalized mass flux stream function 
(Equation \ref{eq_def_Psi}) computed 
based on the single-cell MC profile 
that is inferred without the HK21-type constraint. 
A red (blue) region with dashed (solid) contours 
indicates counterclockwise (clockwise) motion. 
Both hemispheres possess a single cell structure in the radial direction. 
}
\label{fig:4_rev}
\end{center}
\end{figure}
Another feature found in the trade-off planes is 
a nonmonotonous behavior of the residual 
with respect to $\gamma$ for fixed $\alpha$ and $\beta$. 
In particular, 
the residual 
is not monotonic in terms of $\gamma$ 
when $\beta > 10^{-30}$, 
for which we give some explanation. 
We first focus on the dependence of the residual 
on $\beta$ for a fixed $\gamma$. 
It is readily noticed that the residual is monotonically increasing 
in terms of $\beta$. 
As shown later in this section, 
the increasing residual is accompanied by a transition 
in the averaged AM flux by MC 
from the poleward one to equatorward one; 
roughly speaking, $D_{\mathrm{HK1}} \hat{\boldsymbol{u}}_{\mathrm{cand}} \parallel - \boldsymbol{b}$ 
for $\beta < 10^{-30}$ and 
$D_{\mathrm{HK1}} \hat{\boldsymbol{u}}_{\mathrm{cand}} \sim \boldsymbol{b}$ for $\beta > 10^{-30}$ 
(see Section \ref{sec:RLS_HK} 
for the definitions of $D_{\mathrm{HK1}}$ and $\boldsymbol{b}$). 
This trend can be interpreted as evidence that 
an estimated MC profile with stronger equatorward AM transport 
is related to a larger residual. 
Keeping that in mind, let us then focus on 
the dependence of the residual on $\gamma$ 
(see, e.g., $\beta = 10^{-28}$ in Figure \ref{fig:0}). 
When $\gamma$ is small ($\gamma < 10^{-31}$), 
$\beta$ is a dominant trade-off parameter; 
thus, 
the residual increases with the increasing $\gamma$ 
(which is a typical trend 
as discussed in the previous paragraph) 
while achieving the equatorward AM transport by MC 
($D_{\mathrm{HK1}} \hat{\boldsymbol{u}}_{\mathrm{cand}} \sim \boldsymbol{b}$). 
However, $\gamma$ becomes dominant when $\gamma > 10^{-30}$, 
prioritizing minimization of the latitudinal derivative of $\bar{F}_{\mathrm{MC},\theta}(\theta)$, 
eventually leading to a flat solution with $\bar{F}_{\mathrm{MC},\theta}(\theta) \sim 0$ 
in the limit $\gamma \rightarrow \infty$ 
where $D_{\mathrm{HK1}} \hat{\boldsymbol{u}}_{\mathrm{cand}} \sim \boldsymbol{0}$. 
The results show that for $\gamma > 10^{-30}$,
the averaged AM flux by MC transitions from equatorward 
to $0$ 
as $\gamma$ becomes larger, 
corresponding to smaller residuals 
as we discussed the $\beta$ dependence of the residual for a fixed $\gamma$. 


\subsection{Examples of inferred MC profiles} \label{sec:results}
\begin{figure}[t]
\begin{center}
\includegraphics[scale=0.62]{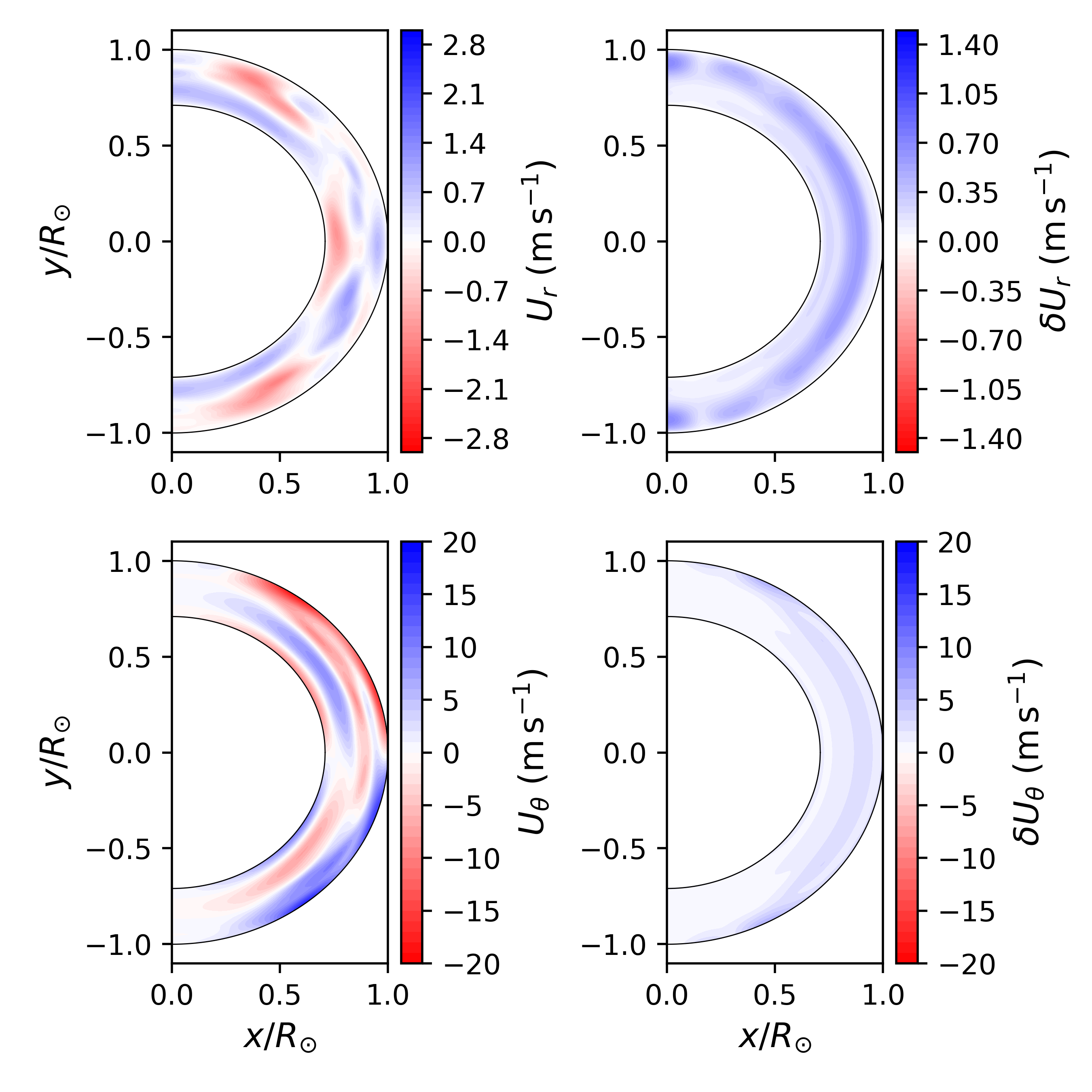}
\caption{\footnotesize Same as Figure \ref{fig:3_rev} 
except that the inversion results were obtained with the constraint on AM transport by MC, 
namely, $\alpha = 10^{-3}$, 
$\beta = 10^{-30}$, and $\gamma = 10^{-31}$, are shown. 
}
\label{fig:5_rev}
\end{center}
\end{figure}
We move on to the reasonable solutions for MC inversion 
chosen based on the procedure described in Section \ref{sec:how2_reason}. 
We first show an example of the results of MC inversion without any constraints 
on AM transport by MC, 
namely, $\beta = \gamma = 0$ (the other trade-off parameter $\alpha$ is set to be $10^{-3}$). 
That corresponds to MC inversion carried out by G20. 
We thus expect the G20 result, and indeed, 
we have confirmed it; that is, 
there is an equatorward flow at the base of the convective zone, 
and the sign of the radial component of the meridional flow field ($U_{r}$) 
is unchanged at most of the latitudes (Figure \ref{fig:3_rev}); 
thus, the single-cell MC profile has been inferred (Figure \ref{fig:4_rev}). 

\begin{figure}[t]
\begin{center}
\includegraphics[scale=0.72]{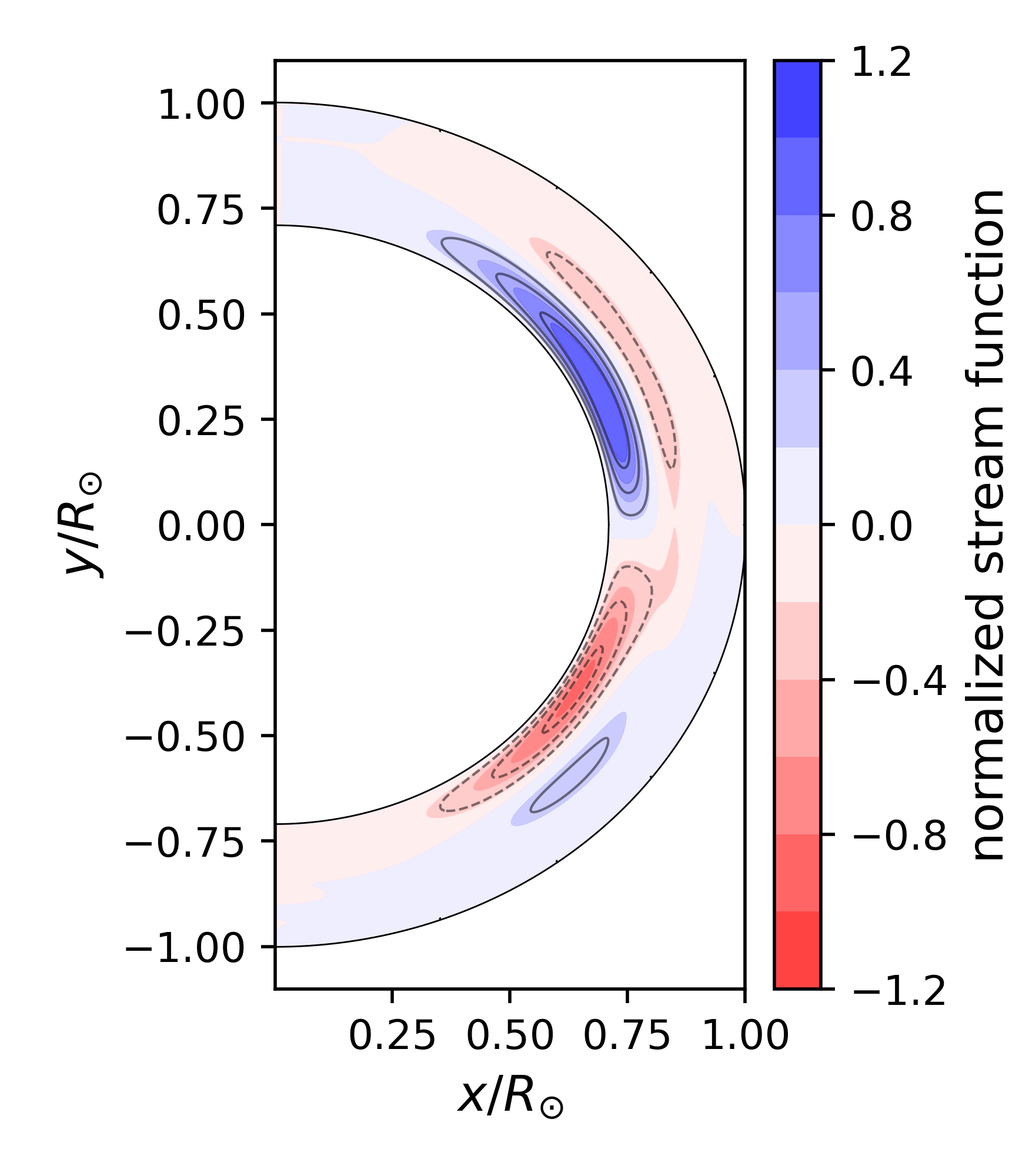}
\caption{\footnotesize Same as Figure \ref{fig:4_rev}, but 
in the case of the MC profile inferred with the constraint on AM transport by MC. 
Each hemisphere has two cells. 
}
\label{fig:6_rev}
\end{center}
\end{figure}
Then, we present inversion results obtained with the HK21-type constraint. 
Figure \ref{fig:5_rev} shows an example of the inferred MC profiles 
for which the trade-off parameters $\alpha = 10^{-3}$, 
$\beta = 10^{-30}$, and $\gamma = 10^{-31}$ are used. 
The prominent features are that there is a poleward flow 
at the base of the convective zone 
and that the radial component of the meridional flow 
changes the sign in the radial direction 
for all the latitudes, 
indicating a double-cell MC profile. 
The double-cell structure can be confirmed as well by 
taking a look at the stream function (Figure \ref{fig:6_rev}) 
that shows 
each hemisphere has two cells in the radial direction, 
although there is a contrast 
between the deeper and shallower cells. 
We would like to emphasize that all the reasonable solutions 
have exhibited the double-cell structure. 
As a whole, MC inversion with the HK21-type constraint 
results in a double-cell MC profile, 
which is different from the G20 result. 

\begin{figure*}[t]
\begin{center}
\includegraphics[scale=0.54]{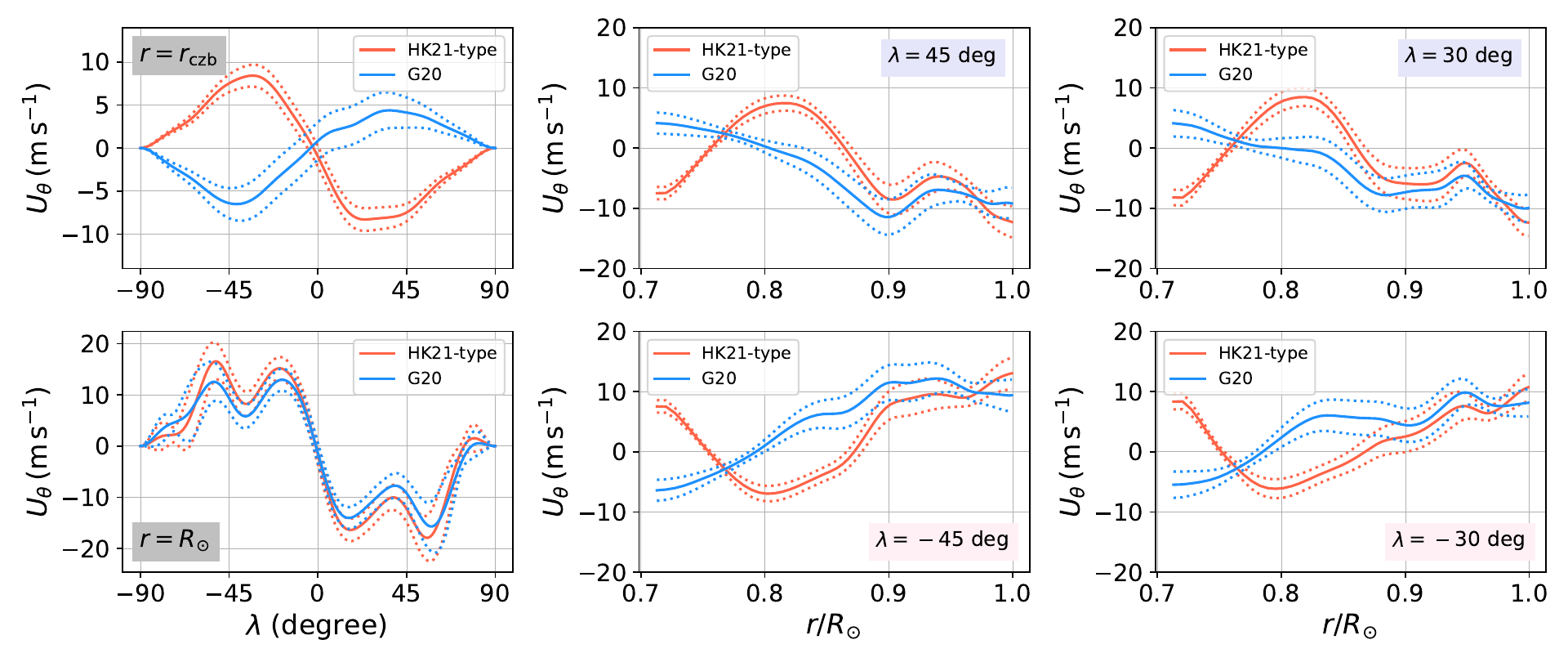}
\caption{\footnotesize Slices of the latitudinal component of the meridional flow field $U_{\theta}$ 
inferred with and without the HK21-type constraint (red and blue, respectively). 
The upper-left and lower-left panels show the slices at 
the bottom of the convective zone ($r = r_{\mathrm{czb}}$) and 
the surface $r = R_{\odot}$, 
respectively. 
The middle and right panels show the slices 
at certain latitudes, namely, 
$\lambda = 45$ degrees (upper-middle), 
$\lambda = -45$ degrees (lower-middle), 
$\lambda = 30$ degrees (upper-right), 
and $\lambda = -30$ degrees (lower-right), respectively. 
Dashed curves indicate $1\sigma$ uncertainties that are evaluated with Equation (\ref{eq_est_err}). 
}
\label{fig:7_rev}
\end{center}
\end{figure*}
\subsection{Comparison of the inferred MC profiles} \label{sec:comparison}
To 
see the difference between 
the inferred single- and double-cell MC profiles 
more quantitatively, 
we present slices of the latitudinal component of 
the inferred meridional flow field ($U_{\theta}$) 
at certain radii ($r=r_{\mathrm{czb}}$ or $R_{\odot}$) 
and latitudes ($|\lambda| = 30$ or $45$ degrees) (Figure \ref{fig:7_rev}). 
It is seen that, in both MC profiles, 
the inferred surface MC velocity ($\sim 10 - 20\, \mathrm{m} \, \mathrm{s}^{-1}$) 
is compatible with the results of surface observations \citep[e.g.][]{1996ApJ...460.1027H}. 
However, we notice a clear difference in 
the flow profiles in a deeper region, 
especially around the base of the convective zone. 
The inferred single-cell MC profile there is characterized by 
an equatorward flow 
whose velocity amplitude is relatively small ($< 5 \, \mathrm{m} \, \mathrm{s}^{-1}$; 
see the upper-left panel in Figure \ref{fig:7_rev}). 
In contrast, in the case of the double-cell MC profile, 
we find a fairly fast poleward flow ($\sim 8-9 \, \mathrm{m}  \, \mathrm{s}^{-1}$) 
whose velocity is even comparable to the surface one at some latitudes. 

Another obvious difference between the single- and double-cell MC profiles is 
the thickness of the subsurface poleward flow 
(see, e.g., the upper-middle panel in Figure \ref{fig:7_rev}). 
The single-cell MC profile exhibits 
the subsurface poleward flow that reaches around $r/R_{\odot} \sim 0.8$, 
whereas that of the double-cell profile is thinner and 
reaches a shallower depth ($r/R_{\odot} > 0.85$). 

\begin{figure}[t]
\begin{center}
\includegraphics[scale=0.55]{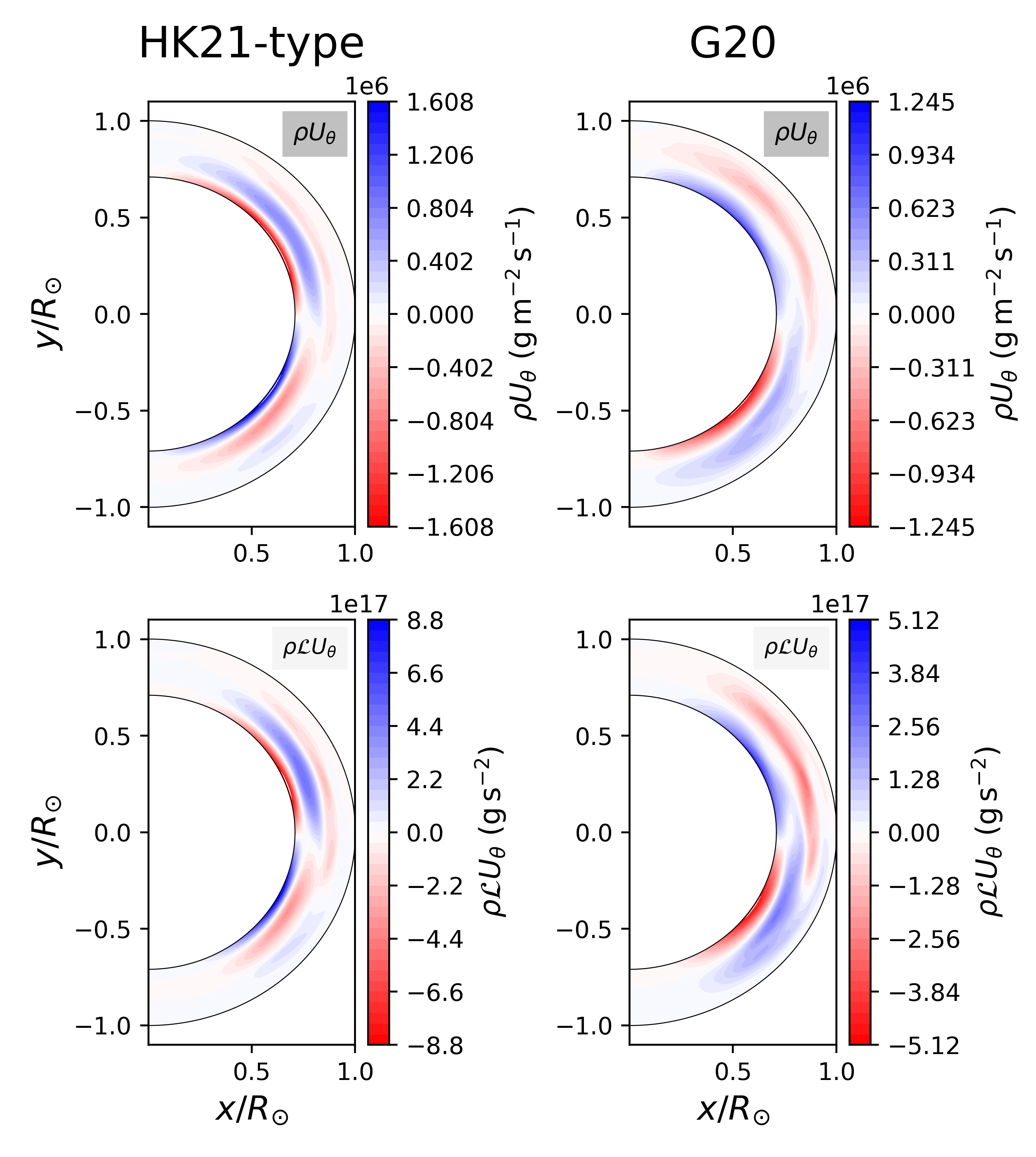}
\caption{\footnotesize Mass flux $\rho U_{\theta}$ (upper panels) and 
AM flux $\rho \mathcal{L} U_{\theta}$ (lower panels) 
calculated with the latitudinal component of 
the inferred meridional flow fields $U_{\theta}$ 
(lower-left panels in Figures \ref{fig:3_rev} and \ref{fig:5_rev}). 
Red (blue) indicates the northward (southward) flux. 
See Appendix \ref{sec:app:c} for how $\rho$ and $\mathcal{L}$ 
were computed. 
}
\label{fig:8_rev}
\end{center}
\end{figure}
The differences between the inferred flow profiles 
are closely related to differences in 
(the latitudinal component of) the mass/AM flux by MC (Figure \ref{fig:8_rev}). 
We first consider the mass flux ($\rho U_\theta$). 
The net mass flux is zero 
at an arbitrary colatitude 
since we assume the mass conservation in this study. 
Therefore, in the case of the single-cell MC profile, 
the equatorward mass flux around the base of the convective zone 
is balanced with the poleward mass flux in the outer envelope 
(see the upper-right panel in Figure \ref{fig:8_rev}). 
This is opposite to the case of the double-cell MC profile 
in which the poleward mass flux in the deep convective zone is 
nearly balanced with the equatorward mass flux in the middle of the convective zone; 
it should be noted that another poleward mass flux in the outermost zone is 
negligible due to the density stratification inside the Sun 
(see the upper-left panel in Figure \ref{fig:8_rev}). 

\begin{figure}[t]
\begin{center}
\includegraphics[scale=0.50]{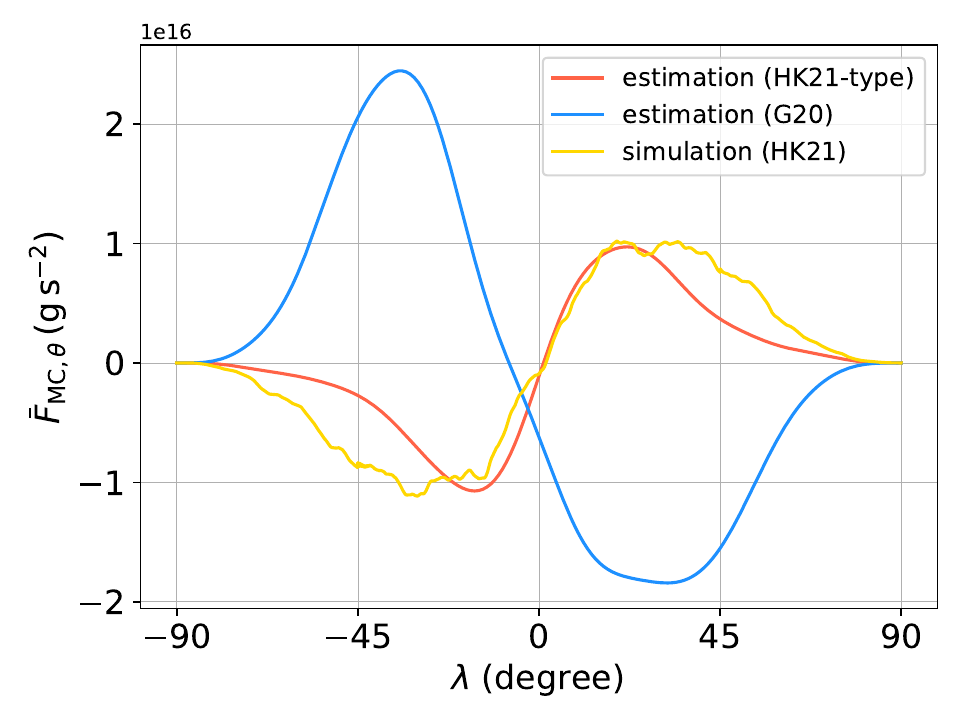}
\caption{\footnotesize Same as Figure \ref{fig:m1} 
except that the averaged AM fluxes 
computed 
based on the inferred single- and double-cell MC profiles 
(Figures \ref{fig:3_rev} and \ref{fig:5_rev}) 
have been included (blue and red, respectively). 
For the HK21-type constraint, 
$\boldsymbol{b} = (5 \times 10^{15}, -5 \times 10^{15})$ 
in units of $\mathrm{g} \, \mathrm{s}^{-2}$ has been adopted 
(see Equation \ref{eq_constrain_AM}) 
based on the numerical results of HK21. 
}
\label{fig:9_rev}
\end{center}
\end{figure}

As for the AM flux ($\rho \mathcal{L} U_\theta$), 
it is apparent that the AM flux profile is similar to 
that of the mass flux ($\rho U_\theta$) (compare the lower and upper panels in Figure \ref{fig:8_rev}). 
However, the net AM flux is not zero 
due to the proportionality $\mathcal{L} \propto r^2$. 
More specifically, 
in the case of the single-cell (double-cell) MC profile, 
the outer poleward (equatorward) AM flux is 
dominant over the inner equatorward (poleward) AM flux 
and the net AM flux by MC is poleward (equatorward) (Figure \ref{fig:9_rev}). 
In fact, it can be shown that the net AM flux by MC 
is always poleward when the MC profile is single-cell, 
which is detailed later in Section \ref{sec:disc1}. 

Finally, we briefly remark on the values of the residual terms 
$|E^{-1/2}(\boldsymbol{\tau} - K \hat{\boldsymbol{u}})|^2 / N_d$ 
in the cases of the single- and double-cell MC profiles. 
As seen in the trade-off plane 
for $\alpha = 10^{-3}$ (Figure \ref{fig:0}), 
the single-cell profile, 
which corresponds to the set of the trade-off parameters $(\beta,\gamma) \rightarrow (0,0)$, 
exhibits a smaller residual ($\sim 0.8288$) 
than the double-cell profile, 
which corresponds to the set of the trade-off parameters $(\beta,\gamma) = (10^{-30},10^{-31})$, 
($\sim 0.8300$). 
There is nevertheless just a $0.1 \, \%$ difference 
between the residuals; thus, 
both profiles explain the data more or less at the same level. 
It should also be instructive to mention that 
the residual is dependent on $\alpha$ 
as well, 
and we have found a number of $\alpha$ 
as reasonable values for the trade-off parameter 
with which 
the residual could vary more than the $\sim 0.1 \, \%$ difference mentioned. 
Therefore, based on the size of the residuals alone, 
it is difficult to favor or disfavor either of the MC profiles at the moment. 
We discuss details on inversion results from a different perspective later in Section \ref{sec:disc3}. 




\section{Discussion} \label{sec:disc}
In this section, we discuss 
the inversion results obtained in Section \ref{sec:result}. 
We firstly assess, from theoretical perspectives, 
the result of the double-cell MC profile 
inferred with the HK21-type constraint 
focusing on AM and magnetic field transport by MC 
(Sections \ref{sec:disc1} and \ref{sec:disc_spot}, respectively). 
We then look for possible MC profiles 
based on an assumption that the Reynolds stress is a dominant source for 
the latitudinal AM transport (RS regime), 
though it is in a rather crude manner (Section \ref{sec:disc2}). 
Finally, all of the MC profiles inferred in this study 
are compared 
(Section \ref{sec:disc3}). 

\subsection{Latitudinal AM transport by single-cell MC is always poleward} \label{sec:disc1}
In Section \ref{sec:result}, 
the double-cell MC profile has been inferred 
based on inversion with a constraint that 
the latitudinal AM transport by MC should be equatorward. 
Actually, 
this is expected. 
An important point is that 
when we assume mass conservation and 
$U_{\theta} = 0$ at the poles, 
which are adopted as the strict constraints in this study 
(see Section \ref{sec:RLS}), 
AM transport by single-cell MC is always poleward. 
The reason is explained in the following paragraphs. 

We here consider a single-cell MC in the northern hemisphere 
where the meridional flow is poleward (equatorward) 
in the outer (inner) convective zone. 
Let us then start with the surface integral of the mass flux at an arbitrary colatitude $\theta$ 
that is defined as 
\begin{equation}
M_{\mathrm{net}} = \int_{r_{\mathrm{czb}}}^{R_{\odot}} \int_{0}^{2 \pi} \rho U_{\theta} r \mathrm{sin} \, \theta \mathrm{d} r \mathrm{d} \phi,  \label{eq_def_M} \nonumber
\end{equation} 
where $\phi$ is the azimuthal angle in the spherical coordinate. 
The surface integral $M_{\mathrm{net}}$ is zero 
because of the assumptions of the mass conservation and $U_{\theta} = 0$ at the poles. 
If we take a point inside the convective zone 
at which $U_{\theta}$ 
changes the sign ($r = r_{0}$), 
the previous argument can be expressed as 
$M_{\mathrm{net}} = M_{\mathrm{in}} + M_{\mathrm{out}} =0$, where 
\begin{equation}
M_{\mathrm{in}} = (2 \pi \mathrm{sin} \, \theta) \int_{r_\mathrm{czb}}^{r_{0}} \rho U_{\theta} r \mathrm{d} r, \label{eq_Min} \nonumber
\end{equation} 
and 
\begin{equation}
M_{\mathrm{out}} = (2 \pi \mathrm{sin} \, \theta) \int_{r_{0}}^{R_{\odot}} \rho U_{\theta} r \mathrm{d} r. \label{eq_Mout} \nonumber
\end{equation} 

Then, we focus on the surface integral of the AM flux at the same colatitude 
that can be expressed as 
$L_{\mathrm{net}} =  L_{\mathrm{in}} + L_{\mathrm{out}}$, where 
\begin{equation}
L_{\mathrm{in}} = (2 \pi \mathrm{sin} \, \theta) \int_{r_\mathrm{czb}}^{r_{0}} \rho \mathcal{L} U_{\theta} r \mathrm{d} r, \label{eq_AMin} \nonumber
\end{equation} 
and 
\begin{equation}
L_{\mathrm{out}} = (2 \pi \mathrm{sin} \, \theta) \int_{r_{0}}^{R_{\odot}} \rho \mathcal{L} U_{\theta} r \mathrm{d} r. \label{eq_AMout} \nonumber
\end{equation} 
When we denote the specific AM at $r=r_{0}$ by $\mathcal{L}_{0}$, 
we have 
\begin{equation}
0 < L_{\mathrm{in}} < \mathcal{L}_{0} \times M_{\mathrm{in}},  \label{eq_AMin_ineq} \nonumber
\end{equation} 
and 
\begin{equation}
L_{\mathrm{out}} < \mathcal{L}_{0} \times M_{\mathrm{out}} < 0.  \label{eq_AMout_ineq} \nonumber
\end{equation} 
This is because $\mathcal{L}$ is 
monotonically increasing in the radial direction \citep{2011ApJ...743...79M}, 
and $\mathcal{L} < \mathcal{L}_{0}$ ($\mathcal{L} > \mathcal{L}_{0}$) 
in the inner (outer) convective zone by definition of $\mathcal{L}_{0}$. 
Note also that $M_{\mathrm{in}}$ ($M_{\mathrm{out}}$) 
is positive (negative) in the northern hemisphere. 
Remembering the definition of $L_{\mathrm{net}}$, 
we have the following relation: 
\begin{equation}
L_{\mathrm{net}} = L_{\mathrm{in}} + L_{\mathrm{out}} 
< \mathcal{L}_{0} (M_{\mathrm{in}} + M_{\mathrm{out}}) = 0,  \label{eq_AM_poleward} \nonumber
\end{equation} 
showing that the net latitudinal AM transport by MC is 
negative in the northern hemisphere. 
Likewise, $L_{\mathrm{net}}$ is positive in the southern hemisphere; that is,
AM transport by MC is always poleward in the case of the single-cell MC profile. 
\begin{figure}[t]
\begin{center}
\includegraphics[scale=0.62]{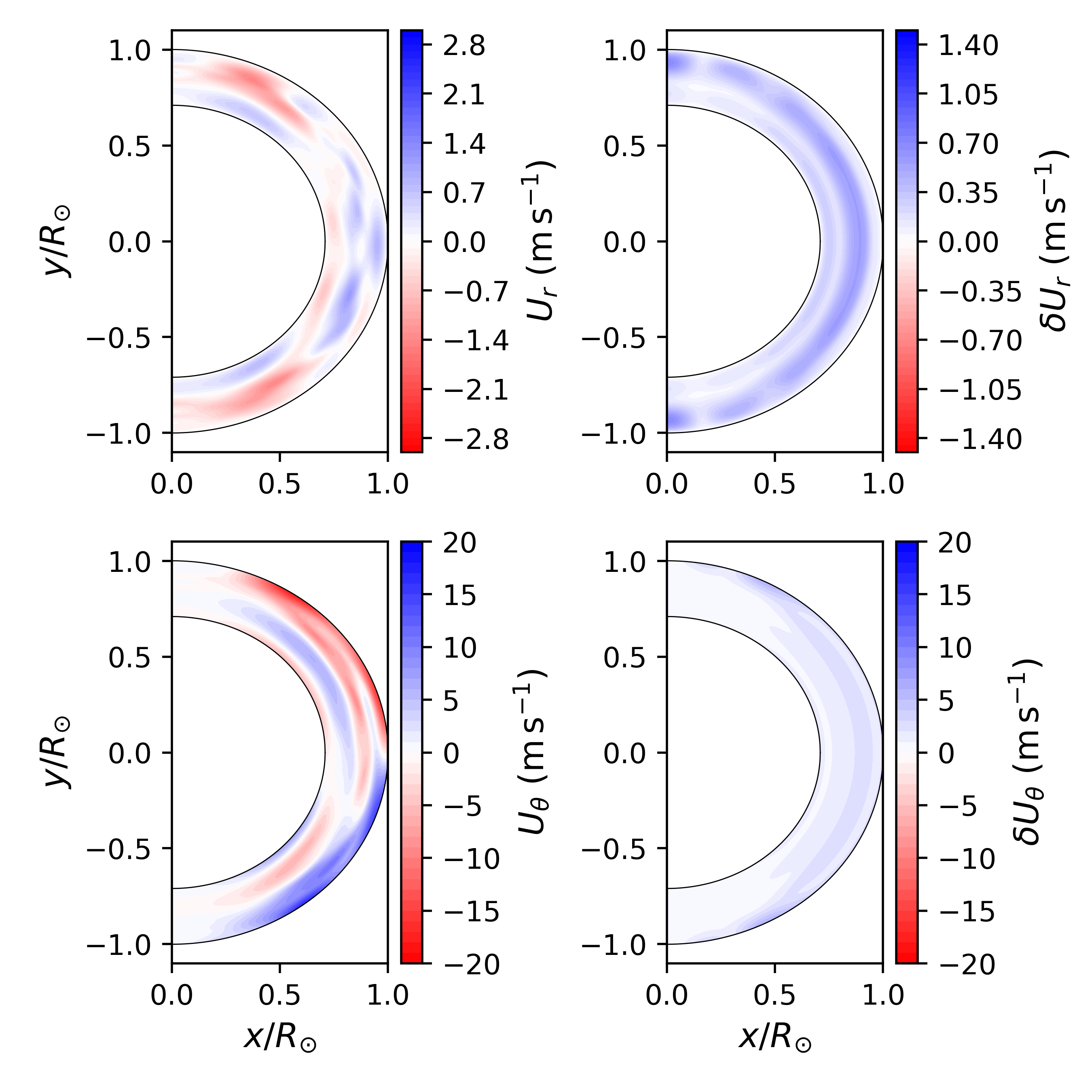}
\caption{\footnotesize Same as Figure \ref{fig:5_rev} 
except that we have weakened the constraint on AM transport by MC 
by requiring a small amount of the equatorward AM transport 
($\sim$ a hundredth of that found in HK21). 
The trade-off parameters $\alpha = 10^{-3}$, $\beta = 10^{-28}$, and 
$\gamma = 10^{-29}$ are used. 
A double-cell MC profile has been inferred 
even with the constraint that 
there is just tiny equatorward AM transport by MC. 
}
\label{fig:10_rev}
\end{center}
\end{figure}

Overall, the poleward AM transport by MC, 
which is inevitable when the MC profile is a single-cell structure, is 
incompatible with the HK21-type constraint 
that AM transport by MC should be equatorward. 
This is the reason why we have not obtained a single-cell MC profile 
via inversion with the HK21-type constraint 
despite the fact that we have used the same dataset as that used in G20 
who have obtained a single-cell MC profile. 
We have confirmed that 
the double-cell MC profile has been inferred 
even when we require a small amount of AM transport by MC 
($\sim$ a hundredth of that found in HK21) (Figure \ref{fig:10_rev}). 
Therefore, 
if the physics in the HK21 regime correctly describes the cause of the solar equator-fast rotation, 
it is plausible that the MC profile is a double-cell structure 
\citep[or it might be multiple-cell that cannot be ruled out because of the low resolution of MC inversion currently available; 
see also discussions in][]{2024ApJ...961...78F}. 

\subsection{How about magnetic field transport by MC?} \label{sec:disc_spot}
We may discuss not only AM transport 
but also magnetic field transport inside the convective zone 
based on the inferred MC profiles. 
For example, 
G20 have found that the single-cell MC can reproduce 
the $11$-year periodic sunspot migration 
by considering the flux transport dynamo 
and assuming that magnetic fields are generated at the base of the convective zone. 
In their framework, the double-cell MC profile cannot explain the equatorward sunspot migration 
since the magnetic fields generated at the base of the convective zone 
are transported toward the poles by the poleward meridional flow there. 

Is the single-cell MC profile preferable to the double-cell profile 
in terms of the magnetic field transport by MC? 
This is not necessarily the case. 
One possible scenario for the equatorward sunspot migration 
by the double-cell MC is that 
magnetic fields are generated in the middle of the convective zone 
where they are transported by 
the equatorward meridional flow \citep[e.g.][]{2013ApJ...762...73N}. 
Alternatively, instead of the flux transport dynamo, 
we can reproduce the equatorward sunspot migration 
by considering the turbulent $\alpha$ effect 
if the sign of $\alpha$ is negative around the tachocline \citep{1955ApJ...122..293P,1975ApJS...29..467Y}. 
It is thus not straightforward to draw some conclusions 
on the morphology of the MC profile 
solely based on magnetic field transport by MC. 


\subsection{What if AM transport by Reynolds stress dominates over AM transport by MC?} \label{sec:disc2}
As described in Section \ref{sec:intro}, 
not only the HK21 regime but also the RS regime 
allows us to reproduce the solar equator-fast rotation \citep[e.g.][hereafter FM15]{2015ApJ...804...67F}. 
In this section, 
in order to discuss MC profiles 
possible in the RS regime, 
we attempt to carry out MC inversion 
with the RS-type constraint. 

In the RS regime, the magnetic field is not dominant in AM redistribution 
and the equator-fast rotation is achieved 
by the turbulent AM transport 
(e.g., FM15). 
It is 
not straightforward to devise constraints on the MC profile 
because 
there is no linear relation between the meridional flow and 
Reynolds stress profile. 
%
%
Therefore, instead of 
physics constraints devised based on the numerical results found in the RS regime, 
we consider qualitative and topological 
constraints that are appropriate for this regime. 
We especially focus on a result that 
the stream function of the mass flux behaves differently 
in the low-latitude region and 
high-latitude region; 
it is 
cylindrical for $|\lambda| < 40$ degrees 
and 
single-cell for $|\lambda| > 40$ degrees, 
where $\lambda$ is the latitude 
(see, e.g., Figures 4 and 6 in FM15). 
We can then separately devise constraints for the low- and high-latitude regions
(more details can be found in Appendix \ref{sec:app:a}). 
Reasonable solutions have been chosen based on the same criteria 
as described in Section \ref{sec:how2_reason}. 
\begin{figure}[t]
\begin{center}
\includegraphics[scale=0.72]{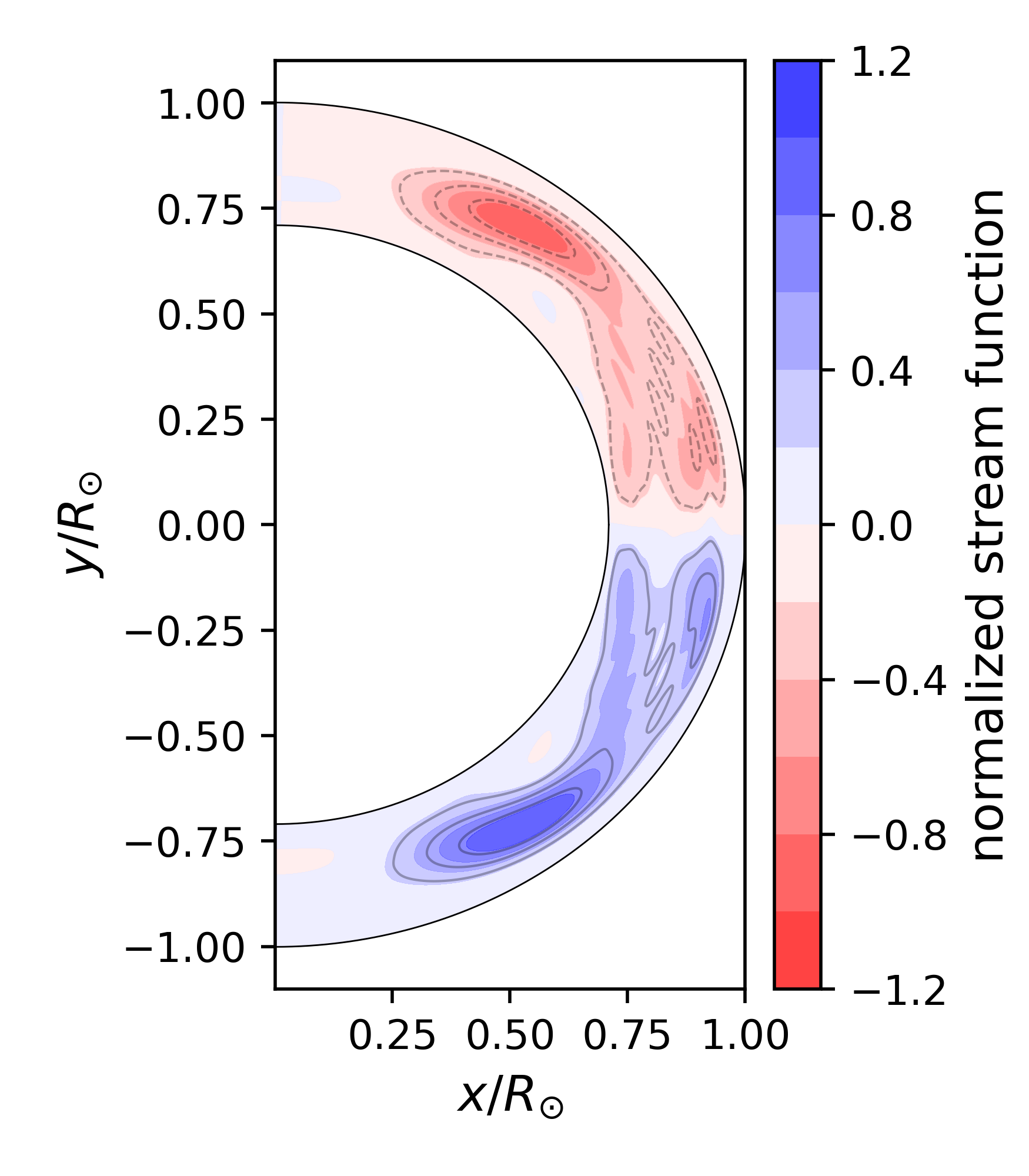}
\caption{\footnotesize Same as Figure \ref{fig:4_rev}, 
but in the case of the MC profile inferred via MC inversion 
with the RS-type constraint 
(see the main text, Appendix \ref{sec:app:a}, and Appendix \ref{sec:app:b} 
for more details). 
The mass flux stream function is characterized by the cylindrical (single-cell) 
profile in the low-(high-)latitude region, 
which is similar to that found in the numerical simulations in the RS regime 
(see, e.g., Figure 6\textit{c} in FM15). 
}
\label{fig:11_rev}
\end{center}
\end{figure}

Figure \ref{fig:11_rev} shows 
the mass flux stream function of the MC profile thus inferred. 
The stream function is a 
single-cell structure in the high-latitude region 
and cylindrical in the low-latitude region, 
resembling the result of numerical simulations in the RS regime 
(see, e.g., Figure 6\textit{c} in FM15). 
More details on the inversion results obtained with the qualitative constraint 
can be found in 
Appendix \ref{sec:app:b}. 

A typical value of the residual terms computed with the reasonable solutions 
found for the RS regime 
is larger than those we have found in Section \ref{sec:result} 
by $\sim 0.5 \, \%$. 
This probably reflects the fact that it is more difficult for us to 
explain the data 
based on an assumption that 
MC profiles should be 
similar to those found in FM15 
compared with the assumptions of the single- or double-cell MC profiles. 


\subsection{Comparison of the inversion results based on averaging kernels} \label{sec:disc3}
In order to 
compare 
the inversion results obtained in this study more carefully, 
we introduce the averaging kernel. 
In the RLS method, 
%
estimates can be expressed as linear combinations of data. 
For instance, 
an estimate of 
the latitudinal component of the meridional flow field 
at a certain target point $(r,\theta) = (r_{\mathrm{t}},\theta_{\mathrm{t}})$ 
is given as: 
\begin{equation}
\hat{U}_{\theta}(r_{\mathrm{t}},\theta_{\mathrm{t}}) =  \sum_{i=1}^{N_{d}} c_{i}(r_{\mathrm{t}},\theta_{\mathrm{t}}) \tau_{i},   \label{eq_est}
\end{equation} 
where $c_{i}$ is the inversion coefficient for the target point that is determined 
by inverting the matrix equation such as Equation 
(\ref{eq_mtx_eq}). 
Substituting Equation (\ref{eq_td_intg}) for Equation (\ref{eq_est}) 
leads to the following expression: 
\begin{eqnarray}
& \, &\hat{U}_{\theta}(r_{\mathrm{t}},\theta_{\mathrm{t}}) =  \nonumber \\ 
& \, & \iint (D_{r}(r,\theta;r_{\mathrm{t}},\theta_{\mathrm{t}}) U_{r}(r,\theta) 
 + D_{\theta}(r,\theta;r_{\mathrm{t}},\theta_{\mathrm{t}}) U_{\theta}(r,\theta) ) \mathrm{d}r \mathrm{d}\theta \nonumber \\ 
& \, & + \sum_{i=1}^{N_{d}} c_{i}(r_{\mathrm{t}},\theta_{\mathrm{t}}) e_{i}, 
 \label{eq_ave_est}
\end{eqnarray} 
where we have the averaging kernel, 
\begin{equation}
D_{\theta}(r,\theta;r_{\mathrm{t}},\theta_{\mathrm{t}}) =  \sum_{i=1}^{N_{d}} c_{i}(r_{\mathrm{t}},\theta_{\mathrm{t}}) 
\mathcal{K}_{i}^{\theta}(r,\theta), \label{eq_ave_ker}
\end{equation} 
and the cross-talk kernel, 
\begin{equation}
D_{r}(r,\theta;r_{\mathrm{t}},\theta_{\mathrm{t}}) =  \sum_{i=1}^{N_{d}} c_{i}(r_{\mathrm{t}},\theta_{\mathrm{t}}) 
\mathcal{K}_{i}^{r}(r,\theta). \label{eq_crs_tlk}
\end{equation} 
The estimate $\hat{U}_{\theta}$ is thus considered as the sum of 
the mean $U_{r}$ averaged by the cross-talk kernel 
and the mean $U_{\theta}$ averaged by the averaging kernel. 
We can evaluate 
the resolution of inversion and contamination in the estimate 
based on the averaging kernel and the cross-talk kernel, respectively. 

\begin{figure*}[t]
\begin{center}
\includegraphics[scale=0.54]{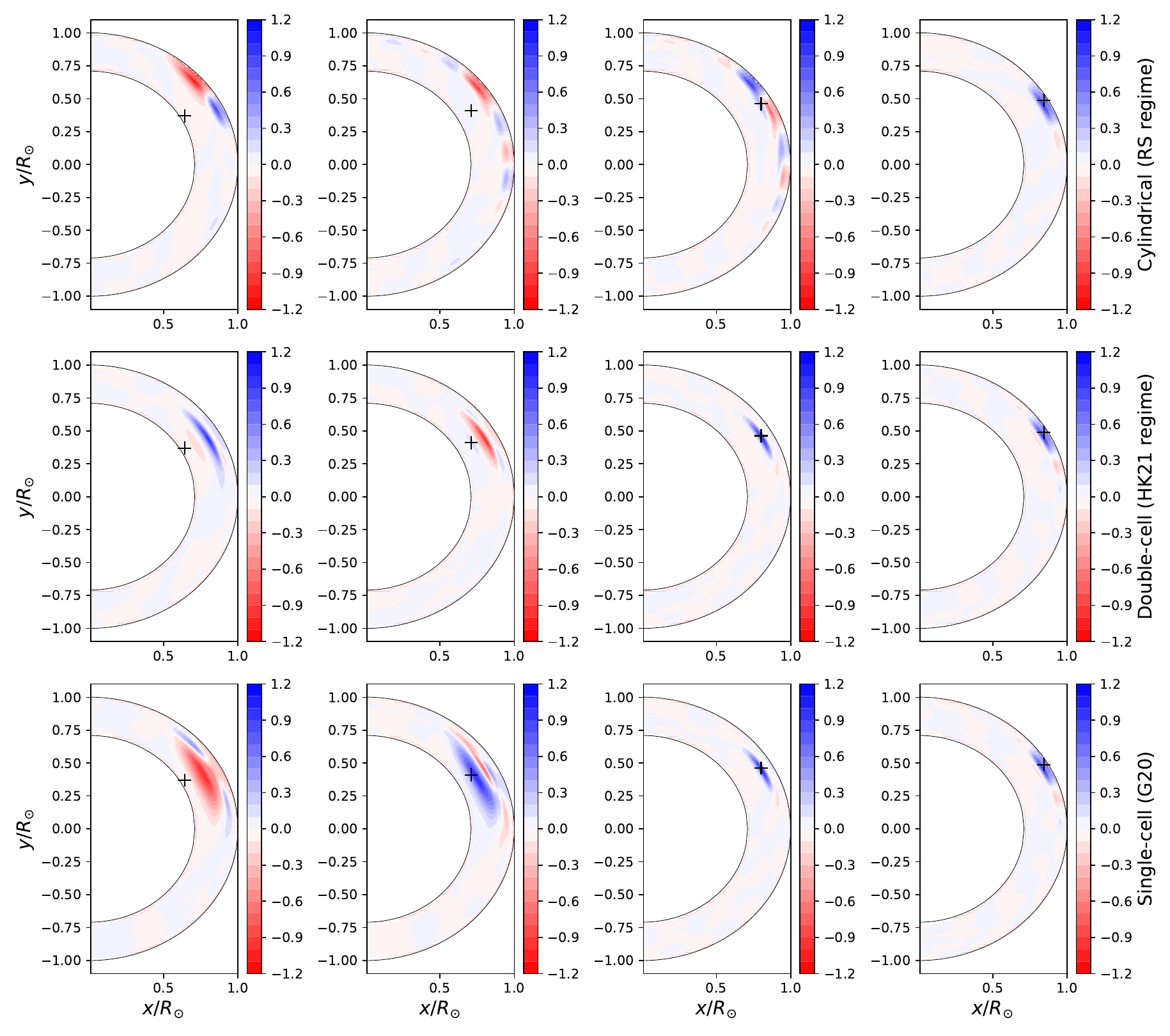}
\caption{\footnotesize Averaging kernels for the inversion results obtained in this study. 
The top, middle, and bottom rows correspond to 
the RS-type constraint, 
HK21-type constraint, and 
the conventional constraints 
(Equations \ref{eq_constrain_mass} and \ref{eq_constrain_conv}), 
respectively. 
Four depths (
$r/R_{\odot} = 0.72$, $0.82$, $0.93$, and $0.98$) 
with a latitude ($\lambda = 30$ degrees) have been chosen as the target points 
that are indicated by black crosses. 
The target depth becomes deeper from right to left. 
Red and blue represent negative and positive values, respectively. 
Well-localized averaging kernels 
exhibit unimodal shapes with positive peaks 
(see, e.g., the second right panels in the middle and bottom rows) 
whereas those with poor resolution exhibit negative values 
and/or oscillatory patterns 
(see, e.g., the second left panel in the top row). 
It should also be noted that we sometimes see well-localized averaging kernels 
but the peak is far from the target point (see, e.g., the leftmost panel in the middle row); hence, the corresponding estimate is thus not considered as 
an appropriate spatial average around the target point. 
Note that the averaging kernels have been normalized for visual aid. \label{fig:12_rev} 
}
\end{center}
\end{figure*}

We show the averaging kernels computed 
for the three different inversion results 
that correspond to the cylindrical, double-cell, and single-cell MC profiles 
(from top to bottom in Figure \ref{fig:12_rev}). 
Four depths ($r / R_{\odot} = 0.72, \, 0.82, \, 0.93,$ and $0.98$) 
at a latitude ($\lambda = 30 $ degrees) are selected as target points. 
Because we have confirmed that 
the cross-talk term (see the first term in Equation (\ref{eq_ave_est})) 
contributes little to the estimate in each case, 
only the averaging kernels are shown in the figure. 

It is evident that, in the case 
that we have obtained the cylindrical MC profile (see the top row in Figure \ref{fig:12_rev}), 
the averaging kernels are well localized 
only in the shallow region ($r/R_{\odot} \sim 0.98$), and 
they 
have exhibited 
some oscillatory patterns 
elongated in the $z$-direction 
that may be caused by the constraint that the stream function should be cylindrical. 
We may find averaging kernels with better resolution 
as we weaken the RS-type constraint by reducing the value of the trade-off parameter, 
though in those cases we cannot reproduce the cylindrical MC profile. 
These properties indicate that 
MC inversion with the RS-type constraint 
is not working well compared with the other two cases 
(the middle and bottom rows in Figure \ref{fig:12_rev}). 

Importantly, when we compare the averaging kernels computed for 
the other two inversion results 
(i.e., the middle and bottom rows in Figure \ref{fig:12_rev}), 
we notice that inversion with the single-cell solution 
performs better than that with the double-cell solution.  
For example, the resolved estimate can be obtained 
even around the middle of the convective zone 
($r/R_{\sun} \sim 0.82$) 
in the case of inferring the single-cell MC profile 
(see the second left panel in the bottom row in Figure \ref{fig:12_rev}). 
The better performance in inversion with the single-cell solution 
is expected because 
the solution has been obtained with the smallest number of the constraints. 
In contrast, in the case with the HK21-type constraint, 
there is a contribution of the extra term 
$\beta D_{\mathrm{HK1}}^{\mathrm{T}} \boldsymbol{b}$ (see the right-hand side of Equation (\ref{eq_mtx_eq_AM})) 
to the estimate 
that becomes 
comparable to half the estimate 
around the base of the convective zone. 
The resolution of averaging kernels 
targeting around $r/R_\odot \sim 0.82$ can be improved 
by reducing the value of the trade-off parameter $\beta$, 
although equatorward AM transport by MC cannot established in such cases, 
as we discussed the resolution of the averaging kernels computed 
for the cylindrical MC profile. 
Thus, the estimates in the deeper region 
are significantly affected by the extra term. 

Then, in terms of how well the localization is done, 
the solution with the single-cell MC profile 
seems to be the most appropriate 
among the inversion results obtained in this study. 
Please note, however, that 
the main purpose of this study is to evaluate how MC inversion is affected 
when we adopt the assumption that 
the HK21 regime correctly describes physics in the solar convective zone. 
We thus cannot conclude that the MC profile is single-cell, 
based on the resolution of the averaging kernels alone. 

Although we have seen in Section \ref{sec:disc1} that the single-cell MC profile 
cannot be inferred via inversion with the HK21-type constraint, 
there is no such restriction 
in the RS regime. 
FM15 discuss the possibility 
that the single-cell MC profile could be realized 
if the Reynolds stress is strong enough 
to overcome the poleward transport AM by MC (see Figure 13 in FM15). 
Thus, as a summary of the inversion results obtained in this study, 
we would like to suggest two possibilities: 
the HK21 regime is correct, and the double-cell MC profile sustains the equator-fast rotation; 
or the RS regime is correct, and the poleward AM transport by the single-cell MC 
is overwhelmed by that of the Reynolds stress. 
One of the keys to narrowing down the possibilities is 
the understanding of the relation between 
the meridional flow field and the Reynolds stress profile 
that would provide us with a new physics constraint on MC inversion 
rather than the crude constraint we attempted in this section. 
We will investigate that point in the forthcoming paper. 

We close this section with a brief comment on the results of MC inversion 
recently carried out by \citet{2023ApJ...954..187H} (hereafter HJ23) 
based on Bayesian statistics. 
With G20's datasets, 
HJ23 have concluded that the MC profile is a single-cell structure 
in both the cycles 23 and 24 (see also Figure 5 in HJ23), 
which is different from our results. 
This may be because how we parameterize the MC profile 
is different from that in HJ23 (see Table 1 in HJ23). 
We also have a difference in the sensitivity kernels; that is,
we have used the kernels computed based on Born approximation 
while 
HJ23 have used ones computed based on ray approximation. 
Some considerations would therefore be required 
to directly compare the results in this study with those in HJ23. 
A comprehensive comparison of possible parameterizations of the MC profile 
would be helpful for selecting the most probable MC profiles, 
which can be conducted via, for example, a Bayesian model comparison \citep{2005blda.book.....G}. 
That is another topic we will work on in the near future. 

\section{Conclusion} \label{sec:conc}
By carrying out inversion of the travel times measured by G20 
with the constraint that 
AM transport by MC should be equatorward (the HK21-type constraint), 
we have inferred a double-cell MC profile. 
As the assumptions of the mass conservation and $U_{\theta}=0$ at the poles 
require that the single-cell MC always transports the AM 
toward the poles, it is plausible that the MC profile is a double-cell structure 
\textit{if AM transport by MC is equatorward inside the Sun, 
as has been found in the HK21 regime}. 


In a rather crude manner, 
MC inversion with the RS-type constraint 
has been carried out, revealing the difficulty in explaining the G20 data 
based on the assumption of a cylindrical mass flux stream function 
that is found in numerical simulations in the RS regime. 
However, 
a comparison of the averaging kernels calculated for the inversion results 
obtained in this study indicates that 
the best localization is achieved for the single-cell MC profile. 
There is thus a dilemma in our inversion, that is,  
putting more priority on the prior 
information (represented by the HK21-type constraint) 
leads us to a double-cell MC profile, 
whereas 
putting more priority on the resolution of the averaging kernels 
leads us to a single-cell MC profile. 
Overall, we have chosen to be conservative 
and not to conclude in this study that the MC profile is a double-cell structure. 


It is theoretically considered that the single-cell MC profile could be established 
if AM transport by the Reynolds stress is fairly strong (RS regime). 
Because the HK21 and RS regimes are currently the only regimes 
in which the solar equator-fast rotation has been reproduced successfully
(and MC inversion with the HK21-type constraint 
has been done in this paper), 
MC inversion with a genuine constraint on the Reynolds stress 
would be desirable for further investigating possible MC profiles in the solar convective zone. 

\section*{Acknowledgements}

We would like to express our gratitude to 
the authors of \citet{2020Sci...368.1469G} 
who have made the datasets 
\dataset[(Open Research Data Repository of the Max Planck Society)]
{https://edmond.mpdl.mpg.de/imeji/collection/0MJjNql7GfpEl5Mb} 
publicly available, 
enabling us to initiate this study. 
This work was supported by 
MEXT as ``Program for Promoting Researches on the Supercomputer Fugaku'' 
(grant numbers: J20HJ00016, JPMXP1020200109, JPMXP1020230406, and JPMXP1020230504), 
JSPS Grant-in-Aid for JSPS Research Fellow Grant Number JP23KJ0300, 
and JSPS KAKENHI grant numbers JP21H04497, JP21H01124, 
JP21H04492, and JP23H01210. 


\bibliography{YH_etal_rev1}{}
\bibliographystyle{aasjournal}



\appendix
\restartappendixnumbering

\section{Constraints on the MC profile devised based on numerical results found in the RS regime \\ 
(the RS-type constraint)} \label{sec:app:a}
%
In this section, we present the details for how we devise the appropriate qualitative constraint for the RS regime 
(see also Section \ref{sec:disc2}). 
We start with the constraint for the low-latitude region 
($|\lambda| < 40$ degrees), 
in which the mass flux stream function is cylindrical 
(see, e.g., Figures 4 and 6 in FM15). 
From the theoretical perspective, 
such cylindrical structures that are aligned 
in the direction of the rotation axis 
are quite common 
in the RS regime 
because of the strong effects of rotation 
on turbulent convection. 

In this study, we define the stream function $\Psi$ 
so that it satisfies the following equation: 
\begin{equation}
\rho \boldsymbol{U} = \nabla \times \biggl ( \frac{\Psi}{r \, \mathrm{sin} \, \theta} \boldsymbol{e}_{\phi} \biggr ), \label{eq_def_Psi}
\end{equation} 
where $\boldsymbol{e}_{\phi}$ is the unit vector in the azimuthal direction. 
Such a stream function always exists because of the mass conservation constraint. 
Whether 
the stream function $\Psi$
is cylindrical or not 
can be quantitatively evaluated by looking at 
the derivative along the polar axis. 
If we define a direction parallel to the polar axis as $z$, 
the derivative along the $z$-axis in the polar coordinate is: 
\begin{equation}
\frac{\partial }{\partial z} 
= 
\mathrm{cos} \, \theta 
\frac{\partial }{\partial r}
- 
\frac{\mathrm{sin} \, \theta}{r} 
\frac{\partial }{\partial \theta}. \label{eq_def_delz}
\end{equation}
Combined with the definition of the stream function (\ref{eq_def_Psi}), 
the $z$-derivative of the stream function can be expressed as: 
\begin{equation}
\frac{\partial \Psi}
{\partial z} 
= 
-\rho r \mathrm{sin} \, \theta \,  
\mathrm{cos} \, \theta \, U_{\theta} 
-\rho r \mathrm{sin}^2 \, \theta \,  
U_{r}  \label{eq_delz}
\end{equation}
that shows a linear relationship between 
the meridional flow field and the first $z$-derivative of the stream function. 
By introducing a matrix $D_{\mathrm{RS}}$, 
we can express the first $z$-derivative of the stream function 
as $D_{\mathrm{RS}} \boldsymbol{u}$. 
Note that we do not consider the outermost region ($r/R_{\odot} > 0.965$) 
because the stream function may not be cylindrical in that region (Hotta in prep.). 
See Appendix \ref{sec:app:c4} for more details 
on how to compute the matrix $D_{\mathrm{RS}}$. 

As for the high-latitude region ($|\lambda| > 40$ degrees) 
where the MC profile is shaped like a single cell (ref), 
we have used the regularization matrix $D$ (see Section \ref{sec:RLS}) 
because it prefers a cellular structure of the MC profile. 
We have just modified the matrix so that 
the meridional flow field in the low-latitude region ($|\lambda| < 40$ degrees), 
except for the outermost region as described in the last paragraph, 
does not contribute 
to regularization (more details can be found in Appendix \ref{sec:app:c2}). 
We denote the modified regularization matrix by $D'$. 

Finally, what we would like to minimize, 
in order to obtain the RLS solution 
with the RS-type constraint, 
ends up with: 
\begin{eqnarray}
X_{\mathrm{RS}}' = | &E&^{-1/2} (\boldsymbol{\tau} - K \boldsymbol{u}) |^{2} 
+ 
\alpha |D' \boldsymbol{u}|^2   
+ 
\boldsymbol{\kappa} \cdot C \boldsymbol{u} 
+ 
\boldsymbol{\mu} \cdot S \boldsymbol{u} \nonumber \\ 
& + & 
\beta |D_{\mathrm{RS}} \boldsymbol{u}|^{2}. 
 \label{eq_S_RLS_FM}
\end{eqnarray}
As shown in Section \ref{sec:RLS_HK}, 
the solution that minimizes the quantity $X_{\mathrm{RS}}'$ 
satisfies the following matrix equation: 
\begin{equation}
\begin{pmatrix}
A'' & C^{\mathrm{T}} & S^{\mathrm{T}} \\
C & O & O \\
S & O & O
\end{pmatrix}
\begin{pmatrix}
\boldsymbol{u} \\
\boldsymbol{\kappa} \\
\boldsymbol{\mu}
\end{pmatrix}
= 
\begin{pmatrix}
K^{\mathrm{T}} E^{-1} \boldsymbol{\tau} \\
\boldsymbol{0} \\
\boldsymbol{0} 
\end{pmatrix}, \label{eq_mtx_eq_FM} 
\end{equation}
where we have introduced a $2N_{j} \times 2 N_j$ matrix $A''$ 
that is defined as: 
\begin{eqnarray}
A'' = A + \beta D_{\mathrm{RS}}^{\mathrm{T}}D_{\mathrm{RS}} \label{eq_submtx_App} 
\end{eqnarray} 
(see Equation (\ref{eq_submtx_A}) for the definition of $A$). 

The solution thus obtained 
is shown in Section \ref{sec:disc} and Appendix \ref{sec:app:b}. 

\section{Example of inferred MC profiles obtained with the RS-type constraint} \label{sec:app:b}
We here show the results of MC inversion carried out with the RS-type constraint. 
We follow the procedures described in Section \ref{sec:how2_reason} 
to determine the values of the trade-off parameters and 
choose reasonable solutions. 

We start with a trade-off plane obtained via inversion. 
\begin{figure}[t]
\begin{center}
\includegraphics[scale=0.52]{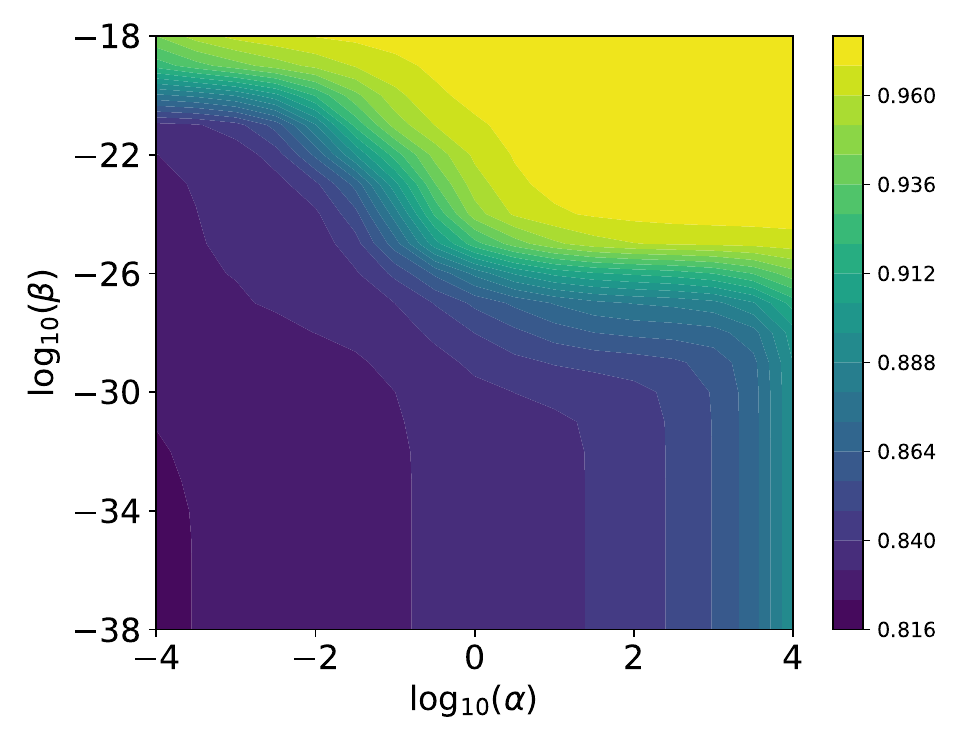}
\caption{\footnotesize 
Trade-off plane in the case of MC inversion with the RS-type constraint. 
The values of the residual term are shown 
as a function of the trade-off parameters $\alpha$ and $\beta$. 
Brighter colors represent larger residual terms. 
The figure shows a trade-off relation between the residual term 
and regularization terms, namely, 
that the residual term becomes larger with the increasing trade-off parameters. 
}
\label{fig:A1}
\end{center}
\end{figure}
The trade-off plane is simpler to understand 
compared with those in the case of the HK21 regime (Figure \ref{fig:0}) 
as the residual changes monotonically against both $\alpha$ and $\beta$ (Figure \ref{fig:A1}). 
This can be explained by the balance between 
minimization of the residual term and regularization 
(represented by the second and fifth terms in Equation (\ref{eq_S_RLS_FM})) 
as we discussed in Section \ref{sec:result}. 
Such a simple trade-off plane may be related to 
the fact that we have imposed two kinds of regularization independently 
for low- and high-latitude regions (see Appendix \ref{sec:app:a}), i.e., 
minimization of the regularization term $\alpha |D' \hat{\boldsymbol{u}}_{\mathrm{cand}}|^{2}$ 
does not conflict with 
minimization of the other regularization term $\beta |D_{\mathrm{FM}} \hat{\boldsymbol{u}}_{\mathrm{cand}}|^{2}$. 

\begin{figure}[t]
\begin{center}
\includegraphics[scale=0.62]{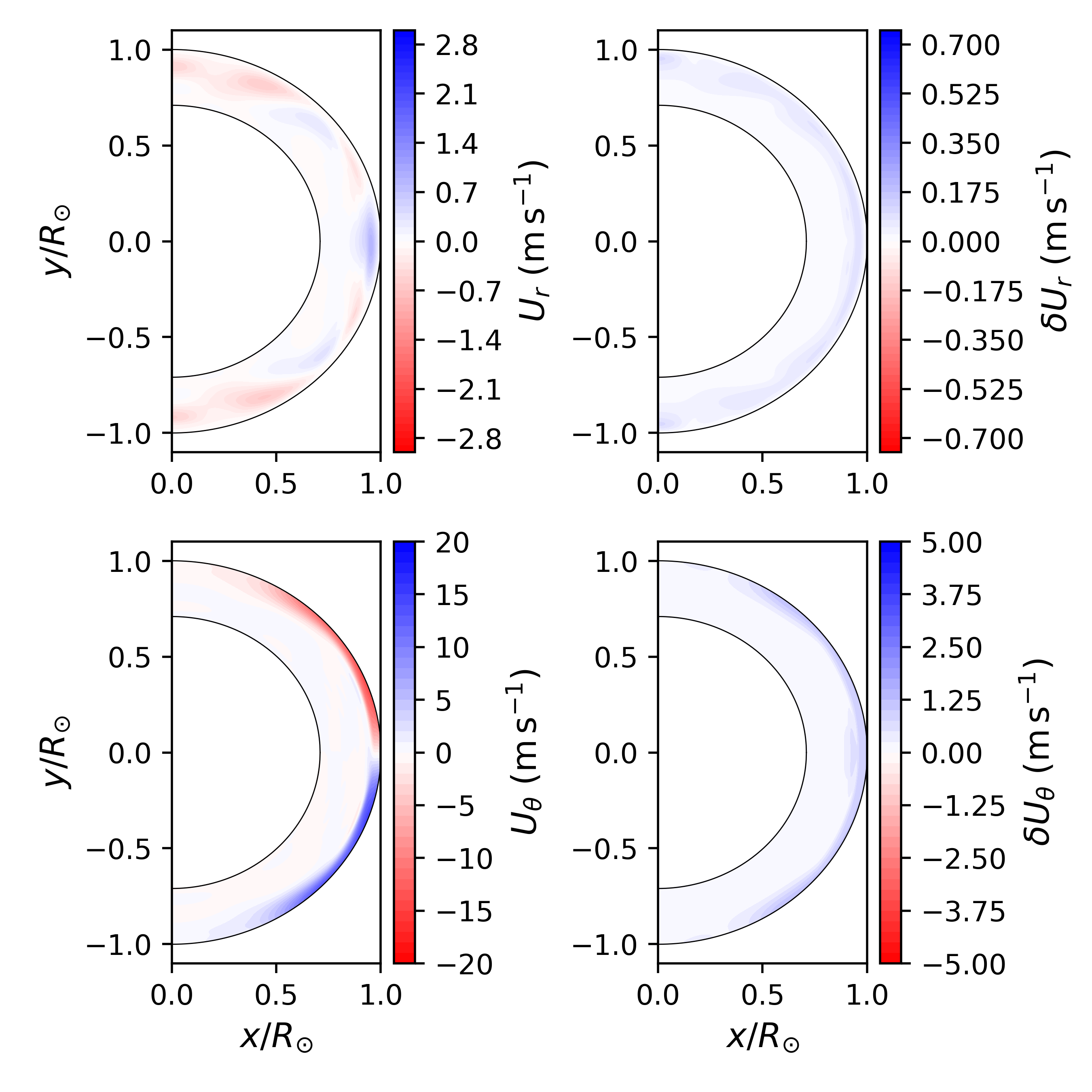}
\caption{\footnotesize 
Radial and latitudinal components of the meridional flow field 
inferred via inversion with the RS-type constraint 
(upper-left and lower-left, respectively). 
The corresponding uncertainties are shown in the right panels. 
The trade-off parameters $\alpha = 10^{-1.6}$ and $\beta = 10^{-14.5}$ are used. 
The results show that the meridional flow velocity is almost zero well inside the convective zone 
($r/R_{\odot} < 0.9$). 
}
\label{fig:A2}
\end{center}
\end{figure}

We then present an example of inferred MC profiles 
that is obtained with the trade-off parameters $(\alpha,\beta) = (10^{-1.6},10^{-14.5})$. 
Figure \ref{fig:A2} shows the radial and latitudinal components of the inferred meridional flow field 
($U_{r}$ and $U_{\theta}$). 
The subsurface MC is characterized by a poleward flow 
as shown in Section \ref{sec:result}, 
and the poleward flow is fairly thin, 
and the amplitude of $U_{\theta}$ is small ($< 3 \, \mathrm{m} \, \mathrm{s}^{-1}$) 
in most of the convective zone ($r/R_{\odot} < 0.9$). 
The radial component $U_{r}$ also exhibits 
a relatively small velocity amplitude $< 0.5 \, \mathrm{m} \, \mathrm{s}^{-1}$ 
inside the low-latitude convective zone 
compared with 
those of the single- or double-cell MC profiles (see Figures \ref{fig:3_rev} and \ref{fig:5_rev}). 
Such an MC profile is required for 
realizing a mass flux stream function with a cylindrical profile in the low-latitude region 
($\lambda < 40$ degrees) 
and a single-cell profile in the high-latitude region ($\lambda > 40$ degrees) 
(see Figure \ref{fig:11_rev} in the main text) 
that is similar to the results found in the RS regime 
(see, e.g., Figure 6\textit{c} in FM15). 

\begin{figure}[t]
\begin{center}
\includegraphics[scale=0.46]{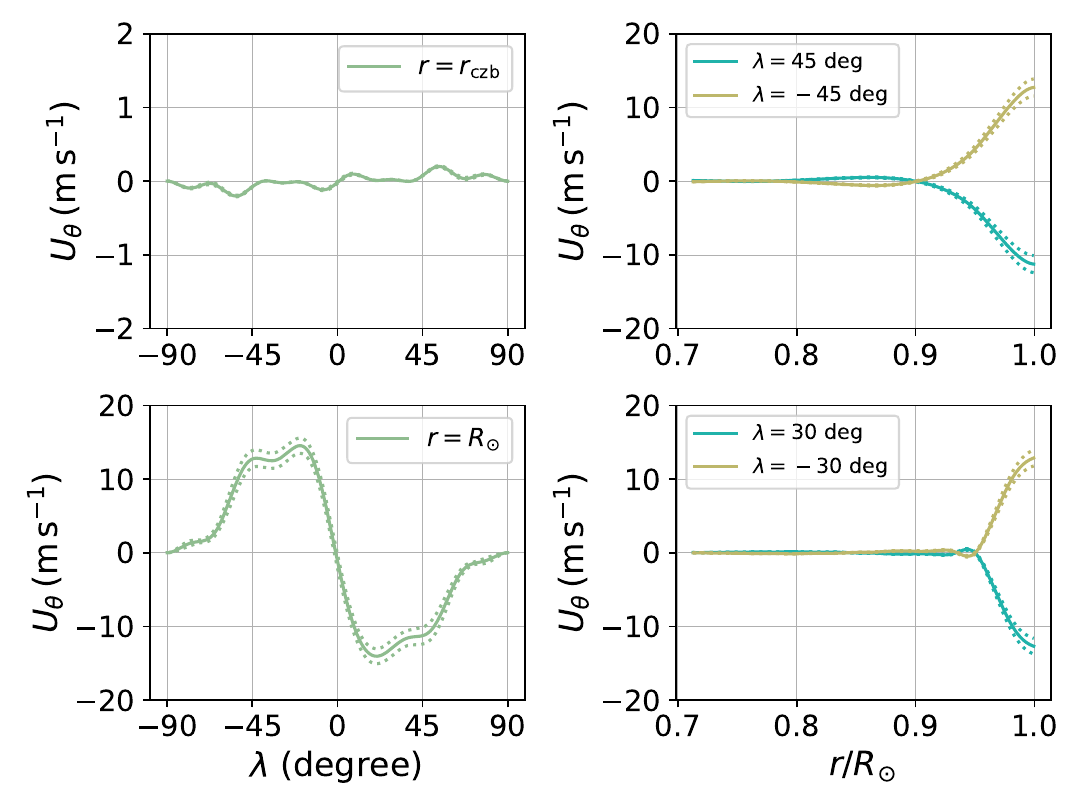}
\caption{\footnotesize Slices of the latitudinal component of the 
inferred cylindrical MC profile. 
The left panels show the slices at certain radii, namely, 
$r=r_{\mathrm{czb}}$ (upper) and $r = R_{\odot}$ (lower), 
and the right panels show those at certain latitudes, 
namely, 
$|\lambda|=45$ degrees (upper) and $|\lambda|=30$ degrees (lower). 
In the right panels, turquoise (sandy yellow) represents the northern (southern) hemisphere. 
}
\label{fig:A3}
\end{center}
\end{figure}

\begin{figure}[t]
\begin{center}
\includegraphics[scale=0.58]{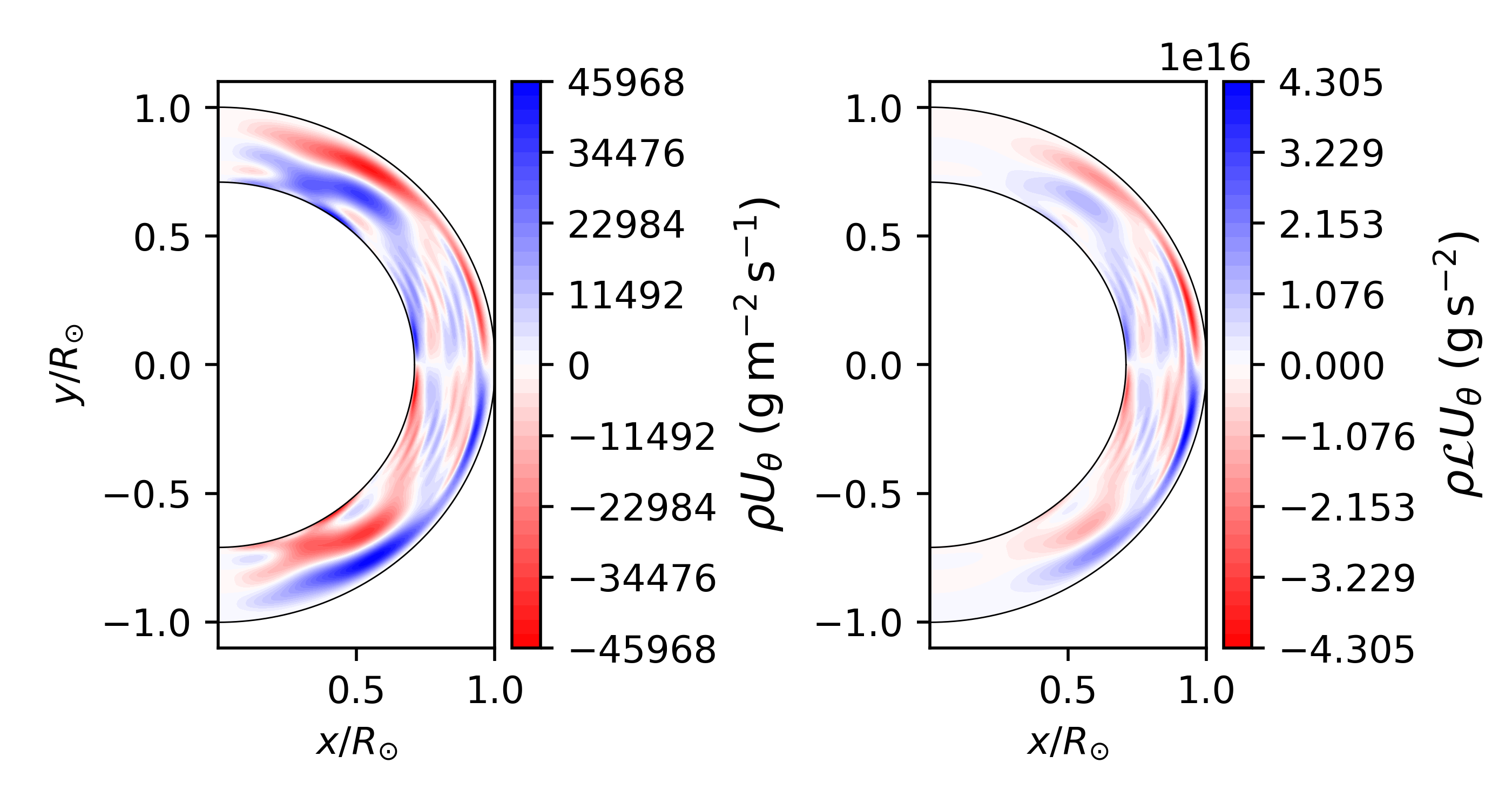}
\caption{\footnotesize Mass flux $\rho U_{\theta}$ (left) 
and AM flux $\rho \mathcal{L} U_{\theta}$ (right) 
computed with the MC profile shown in Figure \ref{fig:A2}. 
The low-latitude region ($|\lambda| < 40$ degrees) is characterized by the cylindrical profile 
in both mass and AM fluxes. 
}
\label{fig:A4}
\end{center}
\end{figure}

\begin{figure}[t]
\begin{center}
\includegraphics[scale=0.52]{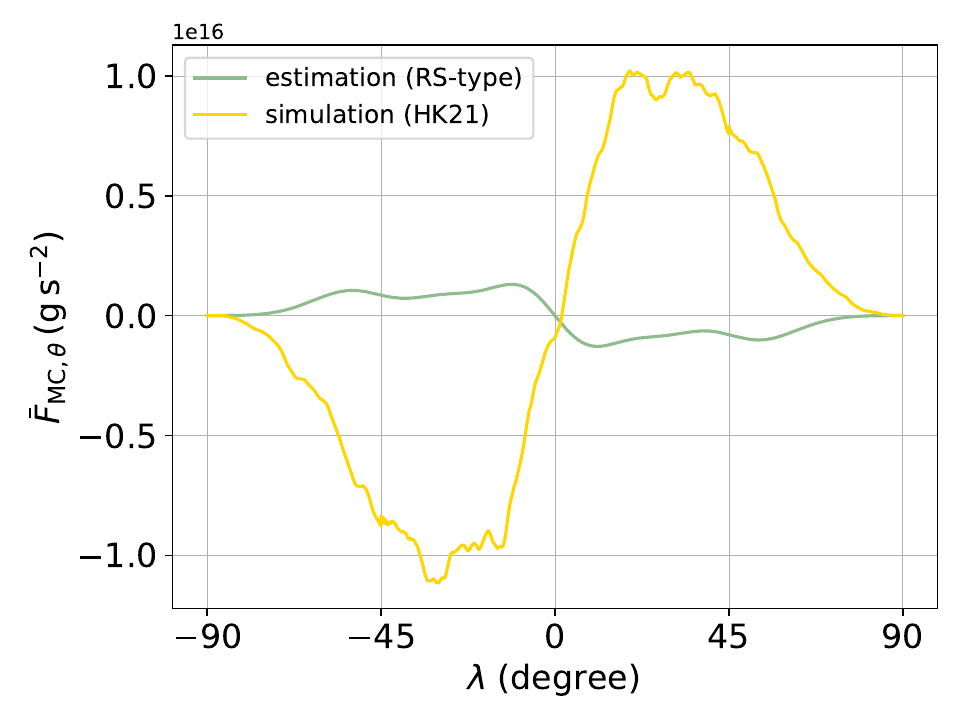}
\caption{\footnotesize Latitudinal AM fluxes by MC that are radially averaged 
$\bar{F}_{\mathrm{MC},\theta}(\theta)$ 
(Equation (\ref{eq_A3_Fmc_th})) as a function of the latitude. 
The averaged AM flux computed with the result of HK21 (yellow) 
indicates the equatorward AM transport 
while that computed with the inferred cylindrical MC profile (green) 
indicates the poleward AM transport. 
It is also seen that AM flux by MC in the case of the cylindrical profile 
contributes less to the total MC transport 
compared with that of the numerical simulation. 
}
\label{fig:A5}
\end{center}
\end{figure}

When we take a look at slices of $U_{\theta}$ (Figure \ref{fig:A3}), 
the difference between the meridional flow field in the low-latitude region 
and that in the high-latitude region is rather obvious. 
While the poleward flow is confined in a narrow layer ($r / R_{\odot} < 0.95$)  
in the low-latitude region, 
it becomes broader in the high-latitude region, 
reaching the depth of $r / R_{\odot} \sim 0.88$ at $\lambda = 45$ degrees. 
We also find an equatorward flow 
below the thick poleward flow 
(see around $0.8 < r/R_{\odot} < 0.9$ of the upper-right panel in Figure \ref{fig:A3}). 
The thicker and faster flows in the high-latitude region lead to 
much more mass flux compared with that in the low-latitude region (left in Figure \ref{fig:A4}). 
Interestingly, the AM fluxes in both latitudes 
are comparable to each other (right in Figure \ref{fig:A4}). 
This is because of the factor $\mathrm{sin} \, \theta$ in $\mathcal{L}$, 
canceling the contrast seen in the mass flux between the high- and low-latitude regions. 
It is also noteworthy to mention that 
the magnitude of the AM flux ($\sim 10^{16} \, \mathrm{g} \, \mathrm{s}^{-2}$) is 
smaller than those of the single- or double-cell MC profiles 
($\sim 10^{17-18}  \, \mathrm{g} \, \mathrm{s}^{-2}$) 
(see Figure \ref{fig:8_rev}), 
highlighting the lower contribution to the net AM flux by MC 
in the case of the MC profile 
inferred with the RS-type constraint (Figure \ref{fig:A5}). 
\section{Notes on the regularization matrices} \label{sec:app:c}
Details on how we have computed the regularization matrices 
are presented in this section. 
After we show a linear relation between 
the meridional flow field $\boldsymbol{U}$ and the expansion coefficient $\boldsymbol{u}$ 
in Appendix \ref{sec:app:c1}, 
specific procedures to compute the regularization matrices, namely, 
$D$ (Appendix \ref{sec:app:c2}), 
$D_{\mathrm{HK1}}$ and $D_{\mathrm{HK2}}$ (Appendix \ref{sec:app:c3}), 
and $D_{\mathrm{RS}}$ (Appendix \ref{sec:app:c4})
are given. 

\subsection{A linear relation between $\boldsymbol{u}$ and $\boldsymbol{U}$} \label{sec:app:c1}
As shown in the main text, 
regularization has been imposed on the meridional flow field $\boldsymbol{U}$, 
but the corresponding regularization terms are expressed 
in terms of the expansion coefficient $\boldsymbol{u}$. 
We thus show a relation between $\boldsymbol{U}$ and $\boldsymbol{u}$ 
in this section. 

We first introduce a $2 N_{D}$-dimensional vector $\boldsymbol{U}_{1\mathrm{d}}$ 
in which $N_{D}$ is the product of the number of mesh points 
in the radial direction $N_{r}$ 
and that in the latitudinal direction $N_{\theta}$. 
In the case of the data provided by G20, 
$N_{r}$ and $N_{\theta}$ are $178$ and $181$, respectively. 
The first $N_{D}$ components in $\boldsymbol{U}_{1\mathrm{d}}$ 
correspond to $U_{r}(r,\theta)$ 
in the following way: 
\begin{equation}
(\boldsymbol{U}_{1\mathrm{d}})_{j'} = U_{r}(r_{k'},\theta_{l'}),  \label{eq_A_U1_Ur}
\end{equation}
where $k'$ runs 
from $1$ to $N_{r}$, 
$l'$ runs from $1$ to $N_{\theta}$, 
and $j' = N_{\theta} \times (k'-1)+ l'$. 
Similarly, the latter $N_{D}$ components in $\boldsymbol{U}_{1\mathrm{d}}$ 
correspond to $U_{\theta}(r,\theta)$ as follows: 
\begin{equation}
(\boldsymbol{U}_{1\mathrm{d}})_{N_{D}+j'} = U_{\theta}(r_{k'},\theta_{l'}),  \label{eq_A_U1_Uth}
\end{equation}
with the same indices used in Equation (\ref{eq_A_U1_Ur}). 

Let us denote the $N_{D}$-dimensional vectors 
(\ref{eq_A_U1_Ur}) and (\ref{eq_A_U1_Uth}) by 
$\boldsymbol{U}_{1\mathrm{d},r}$ and $\boldsymbol{U}_{1\mathrm{d},\theta}$, 
and then Equation (\ref{eq_MCv_expanded}) can be rewritten as 
\begin{equation}
(\boldsymbol{U}_{1\mathrm{d},s})_{j'} = \sum_{j} u_{s,j} Q_{k}(r_{k'}) P_{l}(\mathrm{cos} \, \theta_{l'}),  \label{eq_A_U1_u}
\end{equation}
where 
$k$ runs 
from $1$ to $N_{k}$, 
$l$ runs from $1$ to $N_{l}$, 
$j = N_{l} \times (k-1)+l$, and $s= r$ or $\theta$ (see also Section \ref{sec:data}). 
By introducing a $2N_{D} \times 2N_{j}$ matrix $G$, 
we have a linear relation between $\boldsymbol{U}_{1\mathrm{d}}$ 
and $\boldsymbol{u}$: 
\begin{equation}
\boldsymbol{U}_{1\mathrm{d}} = G \boldsymbol{u},  \label{eq_A_U1_u_mtx}
\end{equation}
where $G$ consists of an $N_{D} \times N_{j}$ matrix $G'$ 
with which 
\begin{equation}
G = 
\begin{pmatrix}
G' & O \\
O & G'
\end{pmatrix}.
 \label{eq_A_G} 
\end{equation}
Elements in the matrix $G'$ are given as 
\begin{equation}
(G')_{j'j} = Q_{k}(r_{k'}) P_{l}(\mathrm{cos} \, \theta_{l'}).  \label{eq_A_Gp}
\end{equation}

\subsection{How to compute the regularization matrix $D$} \label{sec:app:c2}
Following G20, we consider the weighted vorticity of the meridional flow field 
that is defined as $ [\partial_{r}(r U_{\theta}) - 10 \partial_{\theta} U_{r} ] / r $ 
where $\partial_{s}$ stands for the partial derivative 
in terms of the variable $s$. 
We define an $(N_{r} - 1)(N_{\theta}-1) \times 2N_{D}$ matrix 
$D_{\mathrm{v}}$ 
so that the norm $|D_{\mathrm{v}} \boldsymbol{U}_{\mathrm{1d}}|^2$ 
represents the surface integral of the squared weighted vorticity 
over the whole meridional plane at a certain azimuth. 
%
In this paper, 
the forward difference and trapezoidal integration have been adopted, and 
elements of $D_{\mathrm{v}}$ are given in the following way: 
\begin{equation}
(D_{\mathrm{v}})_{j_{0}' j_{1}'} = (D_{\mathrm{v}})_{j_{0}' (j_{2}')} 
= 10 \sqrt{\frac{\Delta r_{k'}}{4(r_{k'+1/2})\Delta \theta_{l'}}},  \label{eq_A2_G20_0}
\end{equation} 
\begin{equation}
(D_{\mathrm{v}})_{j_{0}' (j_{1}'+1)} = (D_{\mathrm{v}})_{j_{0}' (j_{2}'+1)}
= -10 \sqrt{\frac{\Delta r_{k'}}{4(r_{k'+1/2})\Delta \theta_{l'}}},  \label{eq_A2_G20_01}
\end{equation} 
\begin{equation}
(D_{\mathrm{v}})_{j_{0}' j_{3}'} = 
(D_{\mathrm{v}})_{j_{0}' (j_{3}'+1)} = - r_{k'} \sqrt{\frac{\Delta \theta_{l'} }{4(r_{k'+1/2})\Delta r_{k'}}}, 
\label{eq_A2_G20_1}
\end{equation} 
and
\begin{equation}
(D_{\mathrm{v}})_{j_{0}' j_{4}'} = 
(D_{\mathrm{v}})_{j_{0}' (j_{4}'+1)} = r_{k'+1} \sqrt{\frac{\Delta \theta_{l'} }{4(r_{k'+1/2})\Delta r_{k'}}}, 
\label{eq_A2_G20_2}
\end{equation} 
where $\Delta r_{k'} = r_{k'+1} - r_{k'}$, 
$\Delta \theta_{l'} = \theta_{l'+1} - \theta_{l'}$, 
and $r_{k'+1/2} = (r_{k'} + r_{k'+1}) / 2$.  
The indices are defined as below: 
\begin{equation}
j_{0}' = (N_{\theta} - 1) \times (k' - 1) + l', 
\end{equation}
\begin{equation}
j_{1}' = N_{\theta} \times (k' - 1) + l', \label{eq_A2_index_j1}
\end{equation}
\begin{equation}
j_{2}' = j_{1}' + N_{\theta}, 
\end{equation}
\begin{equation}
j_{3}' = j_{1}' + N_{D}, \label{eq_A2_index_1}
\end{equation} 
and 
\begin{equation}
j_{4}' = j_{2}' + N_{D}. 
\label{eq_A2_index_2}
\end{equation} 
Note that $k'$ runs from $n_{\mathrm{czb}}$ to $n_{\mathrm{surf}}-2$ 
with which $r_{n_{\mathrm{czb}}} = r_{\mathrm{czb}}$ and $r_{n_{\mathrm{surf}}} = R_{\odot}$, 
respectively. 
The index $l'$ runs from $1$ to $N_{\theta}-1$. 
The elements other than the ones designated above are zero. 
With the linearity between $\boldsymbol{U}_{\mathrm{1d}}$ 
and $\boldsymbol{u}$ (Equation (\ref{eq_A_U1_u_mtx})), 
the regularization matrix $D$ can be expressed as 
$D = D_{\mathrm{v}} G$. 

As for the modified regularization matrix $D'$ that is used in 
the RS-type constraint 
(see Section \ref{sec:disc2}), 
we prepare an $(N_{r} - 1)(N_{\theta}-1) \times 2N_{D}$ matrix $D_{\mathrm{v}}'$ 
whose 
$j_{0}'$-th row 
is identical to that of $D_{\mathrm{v}}$ 
when a point $(r_{k'},\theta_{l'})$ satisfies either a condition 
$r_{k'}/R_{\odot} > 0.965$ 
or another condition $r_{k'}/R_{\odot} < 0.965$ and $|\pi /2 - \theta_{l'}| > \pi / 9$ 
with $j_{0}' = N_{\theta} \times (k' - 1) + l'$. 
The ranges of the indices are the same as those 
when we compute $D_{\mathrm{v}}$. 
For the other $(p,q)$ elements, 
$(D_{\mathrm{v}}')_{pq} =  0$. 
The matrix $D'$ is then given as $D' = D_{\mathrm{v}}' G$. 

\subsection{How to compute the regularization matrices $D_{\mathrm{HK1}}$ and $D_{\mathrm{HK2}}$} \label{sec:app:c3}
In order to compute the regularization matrices $D_{\mathrm{HK1}}$ and $D_{\mathrm{HK2}}$ 
used in MC inversion with the HK21-type constraint, 
we first consider an $N_{\theta} \times 2 N_{D}$ matrix $D_{\mathrm{HK}}$ 
with which $D_{\mathrm{HK}} \boldsymbol{U}_{\mathrm{1d}}$ represents 
the latitudinal AM flux by MC that is radially averaged, namely, 
\begin{equation}
\bar{F}_{\mathrm{MC},\theta}(\theta) = \frac{\int_{0}^{2 \pi} \int_{r_{\mathrm{czb}}}^{R_{\odot}} F_{\mathrm{MC},\theta}(r,\theta) r \mathrm{sin} \, \theta \mathrm{d}r \mathrm{d} \phi}{\int_{0}^{2 \pi} \int_{r_{\mathrm{czb}}}^{R_{\odot}} r \mathrm{sin} \, \theta \mathrm{d}r \mathrm{d} \phi}. 
\label{eq_A3_Fmc_th}
\end{equation} 
We consider the trapezoidal integration over the radius, 
and elements in the matrix $D_{\mathrm{HK}}$ are given as below: 
\begin{equation}
(D_{\mathrm{HK}})_{l' j_{3}'} 
= \rho_{k'} \mathcal{L}_{k',l'} \times \frac{2 r_{k'} (\Delta r_{k'-1/2})}{(R_{\odot}^2 - r_{\mathrm{czb}}^2)} 
\label{eq_A3_HK21_1}
\end{equation} 
for 
$k' = n_{\mathrm{czb}}+1$ to $n_{\mathrm{surf}}-2$, 
\begin{equation}
(D_{\mathrm{HK}})_{l' j_{3}'} 
= \rho_{k'} \mathcal{L}_{k',l'} \times \frac{r_{k'} \Delta r_{k'}}{(R_{\odot}^2 - r_{\mathrm{czb}}^2)} 
\label{eq_A3_HK21_2}
\end{equation} 
for 
$k' = n_{\mathrm{czb}}$, and 
\begin{equation}
(D_{\mathrm{HK}})_{l' j_{3}'} 
= \rho_{k'} \mathcal{L}_{k',l'} \times \frac{r_{k'} \Delta r_{k'-1}}{(R_{\odot}^2 - r_{\mathrm{czb}}^2)} 
\label{eq_A3_HK21_3}
\end{equation} 
for $ k' = n_{\mathrm{surf}}-1$. 
Note that $l'$ runs from $1$ to $N_{\theta}$. 
The index $j_{3}'$ is defined by Equation (\ref{eq_A2_index_1}). 
The mean $\Delta r_{k'-1/2}$ is defined as $\Delta r_{k'-1/2} = (\Delta r_{k'-1} + \Delta r_{k'})/2$. 
The density at $r = r_{k'}$ is denoted by $ \rho_{k'}$ 
whose value is determined by the standard solar model \citep{1996Sci...272.1286C}.  
The specific AM is $\mathcal{L}_{k',l'} = (r_{k'} \mathrm{sin} \, \theta_{l'})^2 \Omega_{k',l'}$, 
and the averaged angular velocity ($\sim 2 \pi \times 413 \, \mathrm{nHz}$) is used for $\Omega_{k',l'}$. 
The regularization matrix $D_{\mathrm{HK1}}$ represents 
the latitudinal average of $\bar{F}_{\mathrm{MC},\theta}(\theta)$ in each hemisphere. 
We therefore define two matrices for the latitudinal averaging. 
One is an $(N_{\theta} - 1) \times N_{\theta}$ matrix $D_{\mathrm{m}}$ 
defined as follows: 
\begin{equation}
(D_{\mathrm{m}})_{l' l'} = (D_{\mathrm{m}})_{l' (l'+1)} = \frac{1}{2} \label{eq_A3_Dm_1}
\end{equation} 
where $l'$ runs from $1$ to $N_{\theta}-1$. 
The other one is a $2 \times (N_{\theta}-1)$ matrix $D_{\mathrm{int}}$ 
whose elements are 
\begin{equation}
(D_{\mathrm{int}})_{1 l'}  = \frac{\Delta \theta_{l'}}{\pi/2} \label{eq_A3_Dint_1}
\end{equation} 
for $l'=1$ to $N_{\theta}/2$, and 
\begin{equation}
(D_{\mathrm{int}})_{2 l'}  = \frac{\Delta \theta_{l'}}{\pi/2} \label{eq_A3_Dint_2}
\end{equation} 
for $l'=N_{\theta}/2+1$ to $N_{\theta}-1$. 
Using the two matrices $D_{\mathrm{m}}$ and $D_{\mathrm{int}}$ 
in addition to the linear relation (\ref{eq_A_U1_u_mtx}), 
the regularization matrix $D_{\mathrm{HK1}}$ can be expressed as 
$D_{\mathrm{HK1}} = D_{\mathrm{int}} D_{\mathrm{m}} D_{\mathrm{HK}} G$. 

The other regularization matrix $D_{\mathrm{HK2}}$ represents 
the first derivative of $\bar{F}_{\mathrm{MC},\theta}(\theta)$ in terms of the colatitude. 
We have introduced an $(N_{\theta}-1) \times N_{\theta}$ matrix $D_{\theta}$ 
with which the norm $|D_{\theta} \boldsymbol{U}_{\mathrm{1d}}|^2$ is 
the integration of $|\partial_{\theta} \bar{F}_{\mathrm{MC},\theta}|^2$ 
over the colatitude. 
Based on the trapezoidal integration scheme, 
elements in the matrix $D_{\theta}$ are given as 
\begin{equation}
(D_{\theta})_{l' l'} = - (D_{\theta})_{l' (l'+1)} = -\frac{\mathrm{cos}^6 (\theta_{l'+1/2}) + 0.15}{\sqrt{\Delta \theta_{l'}}}, \label{eq_A3_Dth}
\end{equation} 
where $l'$ runs from $1$ to $N_{\theta}-1$. 
The mean colatitude $\theta_{l'+1/2} = (\theta_{l'} + \theta_{l'+1})/2$ 
has been introduced. 
We have an extra term in the numerator 
in addition to the ordinary trapezoidal integration scheme, 
which is considered 
to suppress the rapidly varying $\bar{F}_{\mathrm{MC},\theta}(\theta)$ 
in the high-latitude region. 
We can then express $D_{\mathrm{HK2}}$ as follows: 
$D_{\mathrm{HK2}} = D_{\theta} D_{\mathrm{HK}} G$. 

\subsection{How to compute the regularization matrix $D_{\mathrm{RS}}$} \label{sec:app:c4}
As described in Appendix \ref{sec:app:a}, 
the RS-type constraint 
is represented by using 
the two regularization matrices $D'$ and $D_{\mathrm{RS}}$. 
We have already given how to compute $D'$ 
in Appendix \ref{sec:app:c2}, and we focus on $D_{\mathrm{RS}}$ 
in this section. 
The starting point is the linear relation between 
the first $z$-derivative of the mass flux stream function 
and the meridional flow field (see Equation (\ref{eq_delz})). 
The $z$-derivative of the stream function can then be expressed in a vector form 
as $D_z \boldsymbol{U}_{\mathrm{1d}}$ where 
an $N_{D} \times 2 N_{D}$ matrix $D_{z}$ is defined as follows: 
\begin{equation}
(D_z)_{j_{1}' j_{1}'} 
= - \rho_{k'} r_{k'} \mathrm{sin}^2 \theta_{l'}, \label{eq_A4_RS_1}
\end{equation} 
and 
\begin{equation}
(D_z)_{j_{1}' j_{3}'} 
= - \rho_{k'} r_{k'} \mathrm{sin} \, \theta_{l'} \mathrm{cos} \, \theta_{l'}, \label{eq_A4_RS_2}
\end{equation} 
where $k'$ runs from $1$ to $N_{r}$ and 
$l'$ runs from $1$ to $N_{\theta}$. 
The indices $j_{1}'$ and $j_{3}'$ are defined by Equations 
(\ref{eq_A2_index_j1}) and (\ref{eq_A2_index_1}). 
Because the mass flux stream function is cylindrical 
only in the low-latitude region (see, e.g., Figure 6\textit{c} in FM15), 
we set an entry in $D_{z}$ to be zero 
unless the corresponding point $(r_{k'},\theta_{l'})$ 
falls into a region defined by a condition 
that $r/R_{\odot} < 0.965$ and $|\pi /2 - \theta_{l'}| < \pi / 9$. 
We eventually obtain an expression for the matrix $D_{\mathrm{RS}}$: 
$D_{\mathrm{RS}}=D_z G$. 


\end{document}